\documentclass[%
 reprint,
 amsmath,amssymb,
 aps,
]{revtex4-2}

\usepackage{graphicx}
\usepackage{dcolumn}
\usepackage{bm}
\usepackage{braket}
\usepackage{grffile}

\usepackage{tcolorbox}
\usepackage{mdframed}

\usepackage{amsmath,amssymb}
\usepackage{mathtools}
\usepackage{tikz}
\usetikzlibrary{quantikz}
\usepackage{pgfplots}
\usepackage{subcaption}

\usepackage{cprotect}
\pgfplotsset{compat=1.16}
\usetikzlibrary{positioning}
\usepackage{multirow}

\usepackage{algpseudocode}
\usepackage[boxruled, linesnumbered]{algorithm2e}

\usepackage{hyperref}

\usepackage{tikz}

\begin{document}

\preprint{APS/123-QED}

\title{Simulating excited states of the Lipkin model on a quantum computer}

\author{Manqoba Q. Hlatshwayo$^1$}
\email{manqobaqedindaba.hlatshwayo@wmich.edu}
\author{Yinu Zhang$^1$}
\author{Herlik Wibowo$^2$}
\author{Ryan LaRose$^3$}
\author{Denis Lacroix$^4$}
\author{Elena Litvinova$^{1, 5}$}

\affiliation{%
 $^1$Department of Physics, Western Michigan University, Kalamazoo, MI, 49008, USA\\
  $^2$Institute of Physics, Academia Sinica, Taipei, 11529,  Taiwan \\
 $^3$Department of Computational Mathematics, Science, and Engineering,
Michigan State University, East Lansing, MI 48823, USA \\
$^4$Universit\'{e} Paris-Saclay, CNRS/IN2P3, IJCLab, 91405 Orsay, France\\
 $^5$National Superconducting Cyclotron Laboratory, Michigan State University, East Lansing, MI, 48824, USA}




\date{\today}
\begin{abstract}

We simulate the excited states of the Lipkin model using the recently proposed Quantum Equation of Motion (qEOM) method. The qEOM generalizes the EOM on classical computers and gives access to collective excitations based on quasi-boson operators  $\hat{O}^\dagger_n(\alpha)$ of increasing configuration complexity $\alpha$. We show, in particular, that the accuracy strongly depends on the fermion to qubit encoding. Standard encoding leads to large errors, but  the use of symmetries and the Gray code reduces the quantum resources and improves significantly the results on current noisy quantum devices. With this encoding scheme, we use IBM quantum machines to compute the energy spectrum for a system of $N=2, 3$ and $4$ particles, and compare the accuracy against the exact solution. We found that the results of the approach with $\alpha = 2$, an analog of the second random phase approximation (SRPA), are, in principle, more accurate than with $\alpha = 1$, which corresponds to the random phase approximation (RPA), but the  SRPA is more amenable to noise for large coupling strengths. 
Thus, the proposed scheme shows potential for achieving higher spectroscopic accuracy by implementations with higher configuration complexity, if a proper error mitigation method is applied.

\begin{description}
\item[Keywords] Nuclear Many-Body Problem, Lipkin Model, Quantum Equation of Motion
\end{description}
\end{abstract}

\maketitle


\section{\label{sec:intro} Introduction }

The advent of the digital revolution has brought about a diverse hierarchy of numerical methods for the quantum many-body problem including the mean field theory (MFT) and density functional theory (DFT)~\cite{lacroix2011review, colo2020nuclear}, quantum Monte Carlo (QMC) algorithms~\cite{carlson2015quantum, lynn2019quantum}, machine learning methods~\cite{carleo2017solving}, and others~\cite{schuck2021equation, bogner2013computational}. Whilst these methods have significantly advanced our capabilities to find approximate solutions to the quantum many-body problem, they are all fundamentally limited by the use of classical computers that cannot efficiently simulate quantum physics~\cite{nielsen2000quantum}. Simulating many-body dynamics on quantum computers, which has been proposed over forty years ago to overcome the impediment faced by simulations on classical computers~\cite{feynman1982simulating}, has gained recent attention due to improvements in experimental quantum information processing. Furthermore, simulating nuclear physics on a quantum computer is an emerging area of research addressing both static and dynamic nuclear properties~\cite{zhang2021selected}.  Examples of the former approach, which are based on the Variational Quantum Eigensolver (VQE), include computing the binding energy of light nuclei~\cite{dumitrescu2018cloud,lu2019simulations} and simulation of lattice models~\cite{kokail2019self}. Examples of the latter approach include a quantum algorithm for the linear response theory~\cite{roggero2019dynamic}, the time-evolution of a nuclear many-body system~\cite{guzman2021calculation, guzman2022accessing, roggero2020preparation}, and simulation of non-Abelian gauge theories with optical lattices~\cite{tagliacozzo2013simulation}. Other efforts in the field address efficient state preparation schemes~\cite{roggero2020preparation, lacroix2020symmetry} and analysis of nuclear structure using entanglement~\cite{robin2021entanglement}. \

Modern nuclear experiments provide high-resolution data in the keV range~\cite{NNDC}, while for theoretical calculations~\cite{LitvinovaSchuck2019} it is still challenging to reproduce excitation spectra of medium-mass and heavy nuclei with the accuracy of $\sim 100$ keV (which is $\sim 10\%$ of the atomic nuclear energy scale). Although many approximation techniques that go beyond the MFT and Random Phase Approximation (RPA) have been developed in the last few decades, we still do not have a unified method that achieves spectroscopic accuracy for such nuclear systems. Meanwhile, very accurate nuclear structure input is needed by the applications at the frontiers of nuclear research, such as the astrophysical simulations of kilonova~\cite{cowan2021origin} and supernova~\cite{janka2007theory} as well as the searches beyond the standard model in the nuclear domain~\cite{suhonen1998weak, yamanaka2017review}. We note from quantum chemistry calculations on classical computers, that chemical accuracy (i.e, the errors less than 1 kcal/mol=0.043 eV which are $\sim1\%$ of the probed energy scale) can be achieved using the canonical coupled cluster (CC) expansion truncated at the second order in the electronic excitation operator and including an approximate treatment of the triple excitations CCSD(T), where S stands for single, D for double, and (T) for non-iterative triple~\cite{raghavachari1989fifth, bartlett2007coupled}. This indicates that three-particle-three-hole ($3p3h \coloneqq \alpha = 3$) configuration complexity is sufficient for accurate quantum chemistry calculations, which can be, alternatively to CC, performed within the linear response theory or the equation of motion technique.  However, nuclear calculations with the same ($\alpha =3$) configuration complexity  do not always lead to spectroscopically accurate results~\cite{Ponomarev1999,Savran2011,LoIudice2012,Hagen2014,LitvinovaSchuck2019,Morris2018,Lenske:2019ubp,Sun2021}, because the interaction between nucleons in nuclei is (i) much stronger and (ii) only known with limited accuracy. \

The most general equation of motion (EOM) framework for the quantum $N$-body problem requires $N$ coupled EOMs or, equivalently, the excitation operators of complexity $\alpha = N$ for the exact solution, while the most advanced classical computation of medium-mass and heavy nuclei ($N\sim 100$) hardly reaches the complexity of $\alpha = 3$.
Although the associated accuracy is quite good compared to the accuracy of RPA and many of the gross and even fine features of the nuclear spectra can be captured quite reasonably, in many cases this accuracy is insufficient. Furthermore, there is no firm criterion for the complexity needed for various  nuclear spectral calculations. Therefore, one of the goals for this work is to investigate how the configuration complexity of the many-body states within the EOM framework correlates with the accuracy of the resulting spectra when simulated on a quantum computer. Since direct large-$\alpha$ calculations are still prohibitive for classical computing, it is highly desirable to develop (as an alternative) an efficient quantum algorithm which can be implemented on available Noisy Intermediate-Scale Quantum (NISQ) ~\cite{preskill2018quantum} computers to guide future nuclear structure calculations with configuration complexity $\alpha > 3$.  \ 

In this work, we eliminate the issue of unknown nuclear forces by considering a model Hamiltonian with a tunable two-body interaction. The exactly solvable Hamiltonians represent an ideal playground for such studies as they offer firm benchmarks of the accuracy of the approximate methods. Inspired by quantum chemistry simulations on NISQ computers, we use the recently proposed classical-quantum algorithm, the Quantum Equation of Motion (qEOM)~\cite{Ollitrault_2020}, which is an extension of VQE for computing excitation energies. We simulate the excited states and energies of the Lipkin-Meshkov-Glick (LMG) Hamiltonian~\cite{lipkin1965validity} with configuration complexity $\alpha =1$ (analog of RPA) and $\alpha =2 $ (second RPA (SRPA)). We run the qEOM algorithm on IBM quantum computers for LMG systems with small number of particles $N=2, 3,$ and $4$, and then compare our results with the exact solution, classical Hartree-Fock and RPA solutions. Part of this work builds upon the work done in Ref. ~\cite{cervia2021lipkin}, where the authors introduced an encoding scheme for the Lipkin model and simulated its ground state energy on a quantum computer for a system of $N=2$ particles. We propose a more efficient encoding scheme and simulate both ground and excited state energies for systems of up to $N=4$ particles on a quantum computer. 

The paper is organized as follows: Section \ref{sec:background} gives the background of the Quantum Equation of Motion, the LMG model, and previously used encoding schemes ~\cite{cervia2021lipkin} for the LMG Hamiltonian. In Section \ref{sec:efficient} we present our new efficient encoding scheme which exploits symmetries in the Hamiltonian and employs the Gray encoding to minimize the required quantum resources. The simulation results are shown in Sections \ref{sec:results}, and the summary and outlook are given in Section \ref{sec:summary}.

\section{\label{sec:background}Background}

\subsection{\label{sec:qEOM}Quantum Equation of Motion}

First proposed by Rowe in 1968~\cite{rowe1968equations}, the \emph{Equation of Motion} (EOM) is a framework for computing excitation properties of quantum many-body systems. Given the many-body ground state $\ket{gs}$, we construct an excitation operator $\hat{O}^\dagger_n$ that generates all the excited states  $\ket{n}$ from the ground-state, such that
\begin{equation}\label{exc_op}
	\hat{O}^\dagger_n \ket{gs} = \ket{n} \qquad \text{and} \qquad \hat{O}_n \ket{gs} = 0. 
\end{equation}
The EOM prescription for constructing $\hat{O}^\dagger_n$ involves four steps. First, estimate the ground state $\ket{gs}$ using a suitable approximation like the uncorrelated Hartree-Fock (HF) or the correlated RPA ground state. Second, express $\hat{O}^\dagger_n$ as a linear combination of basis excitation operators with variable expansion coefficients given by 
\begin{equation}\label{eq_excOp}
	\hat{O}^\dagger_n = \sum_{\alpha} \sum_{\mu_{\alpha}}    \left[ X^{\alpha}_{\mu_{\alpha}} (n) \hat{K}^{\alpha}_{\mu_{\alpha}} - Y^{\alpha}_{\mu_{\alpha}} (n) \left(\hat{K}^{\alpha}_{\mu_{\alpha}}\right)^\dagger \right], 
\end{equation}
where $\alpha$ is the degree of configuration complexity and $\mu_{\alpha}$ is the collective index associated with the single-particle (sp) states. A commonly used basis for the excitation operator is the fermionic particle creation and annihilation operator, in which we can write $ \hat{K}^{1}_{\mu_{1}} = a^{\dagger}_{i}a_{j^\prime}$ for $\alpha=1$ (RPA) and $ \hat{K}^{2}_{\mu_{2}} = a^{\dagger}_{i} a^{\dagger}_{j} a_{j^\prime} a_{i^\prime}$ for $\alpha=2$ (second RPA). Note that the indices without (with) the prime represent the particle (hole) states. Third, use Eq. (\ref{exc_op}) and the Schr\"{o}dinger's equation to get the excitation energy above the ground state ($E_{n0} = E_n - E_0$) ~\cite{ring2004nuclear} given by
\begin{equation}
	E_{n0} = \frac{ \braket{\left[ \hat{O}_n, \left[ \hat{H}, \hat{O}^\dagger_n \right]  \right]}  }{ \braket{\left[ \hat{O}_n, \hat{O}^\dagger_n \right]} }, 
\end{equation}
where $\braket{.}$ is a shorthand notation for $\bra{gs}.\ket{gs}$. Fourth, take the variation $\delta (E_{n0}) = 0$ in the parameter space spanned by the coefficients of Eq. (\ref{eq_excOp}) which leads~\cite{Ollitrault_2020} to the generalized eigenvalue equation (GEE)
\begin{equation}\label{eq_gev}
	\begin{pmatrix} \mathcal{A} &  \mathcal{B} \\ \mathcal{B}^{*} & \mathcal{A}^{*} \end{pmatrix}
	\begin{pmatrix} X^{n} \\ Y^{n} \end{pmatrix}
	=
	E_{n0}
	\begin{pmatrix} \mathcal{C} & \mathcal{D} \\ -\mathcal{D}^{*} & -\mathcal{C}^{*} \end{pmatrix}
	\begin{pmatrix} X^{n} \\ Y^{n} \end{pmatrix},
\end{equation}
where the matrices $\mathcal{A}, \mathcal{B}, \mathcal{C}, \text{and} \; \mathcal{D}$ are given by 
\begin{eqnarray}
	 \mathcal{A}_{\mu_\alpha \nu_\beta} &=& \braket{ \left[ \left(\hat{K}^{\alpha}_{\mu_{\alpha}}\right)^\dagger, \left[ \hat{H}, \hat{K}^{\beta}_{\nu_{\beta}} \right]  \right] }, 	\\ 
	 \mathcal{B}_{\mu_\alpha \nu_\beta} &=& - \braket{ \left[ \left(\hat{K}^{\alpha}_{\mu_{\alpha}}\right)^\dagger, \left[ \hat{H}, \left(\hat{K}^{\beta}_{\nu_{\beta}} \right)^\dagger\right]  \right] }, \\ 
	 \mathcal{C}_{\mu_\alpha \nu_\beta} &=& \braket{ \left[ \left(\hat{K}^{\alpha}_{\mu_{\alpha}}\right)^\dagger, \hat{K}^{\beta}_{\nu_{\beta}}  \right]}, 	\\ 
	 \mathcal{D}_{\mu_\alpha \nu_\beta} &=& - \braket{ \left[ \left(\hat{K}^{\alpha}_{\mu_{\alpha}}\right)^\dagger,  \left(\hat{K}^{\beta}_{\nu_{\beta}} \right)^\dagger  \right] }. 
\end{eqnarray}

As an example, we evaluate the matrices $\mathcal{C}$ and $\mathcal{D}$ for excitation configurations with $\alpha =1$. First, note that in the particle-hole ($ph$) representation, the fermionic anti-commutation relations are given by
\begin{equation}\label{eq_ferm_anti_com}
	\begin{split}
		& \lbrace \hat{a}_{\mu} , \hat{a}^\dagger_{\nu} \rbrace = 
		\begin{cases}
			0 \, \, \,  \text{if} \, \, \, \mu\nu = ph \, \, \, \text{or} \, \, \, hp \\
			\delta_{\mu\nu} \, \, \,  \text{if} \, \, \, \mu\nu = pp \, \, \, \text{or} \, \, \, hh 
		\end{cases} \\
		&\lbrace \hat{a}_{\mu} , \hat{a}_{\nu} \rbrace = \lbrace \hat{a}^\dagger_{\mu} , \hat{a}^\dagger_{\nu} \rbrace = 0\\
	\end{split}.
\end{equation}
The excitation operator in the RPA can be then explicitly written as
\begin{equation}\label{eq_excitation_op}
	\hat{O}^{\dagger}_n = \sum_{ij^\prime} \left[ X^{(1)}_{ij^\prime}(n) a^{\dagger}_{i}a_{j^\prime} - Y^{(1)}_{ij^\prime}(n) a^{\dagger}_{j^\prime}a_{i} \right].
\end{equation} 
Using Eq. (\ref{eq_ferm_anti_com}) we first evaluate the simpler matrix $\mathcal{D}$ to get
\begin{equation}
	\begin{split}
		\mathcal{D}_{mi^\prime kj^\prime} &= - \braket{ \left[  \hat{a}^\dagger_{i^\prime} \hat{a}_{m}  , \hat{a}^\dagger_{j^\prime} \hat{a}_{k}   \right] } \\
		&=- \braket{ \left(  \hat{a}^\dagger_{i^\prime} \hat{a}_{m}\hat{a}^\dagger_{j^\prime} \hat{a}_{k} -  \hat{a}^\dagger_{j^\prime} \hat{a}_{k}  \hat{a}^\dagger_{i^\prime}   \hat{a}_{m}  \right) }\\
		&=- \braket{ \left( \hat{a}^\dagger_{i^\prime}  \hat{a}^\dagger_{j^\prime}  \hat{a}_{k} \hat{a}_{m} - \hat{a}^\dagger_{i^\prime} \hat{a}^\dagger_{j^\prime}  \hat{a}_{k}  \hat{a}_{m}  \right) }\\
		&= 0.\\
	\end{split}
\end{equation}
A similar calculation can be done for the matrix $\mathcal{C}$ and its commutator yields
\begin{align*}
	&\left[  \hat{a}^\dagger_{i^\prime} \hat{a}_{m}  , \hat{a}^\dagger_{k} \hat{a}_{j^\prime}   \right] =  \hat{a}^\dagger_{i^\prime} \hat{a}_{m} \hat{a}^\dagger_{k} \hat{a}_{j^\prime} - \hat{a}^\dagger_{k} \hat{a}_{j^\prime} \hat{a}^\dagger_{i^\prime} \hat{a}_{m} \\
	&=  \hat{a}^\dagger_{i^\prime} \left( \delta_{mk} -  \hat{a}^\dagger_{k}\hat{a}_{m} \right)  \hat{a}_{j^\prime} - \hat{a}^\dagger_{k} \left(\delta_{i^\prime j^\prime}  - \hat{a}^\dagger_{i^\prime} \hat{a}_{j^\prime}  \right)  \hat{a}_{m} \\
	&= \delta_{mk}  \hat{a}^\dagger_{i^\prime} \hat{a}_{j^\prime} - \delta_{i^\prime j^\prime}  \hat{a}^\dagger_{k} \hat{a}_{m},
\end{align*}
hence can be written as
\begin{equation}\label{eq_C}
	\begin{split}
		\mathcal{C}_{mi^\prime kj^\prime} &=  \braket{ \left[  \hat{a}^\dagger_{i^\prime} \hat{a}_{m}  , \hat{a}^\dagger_{k} \hat{a}_{j^\prime }   \right] } \\
		&= \braket{ \left( \delta_{mk}  \hat{a}^\dagger_{i^\prime} \hat{a}_{j^\prime} - \delta_{i^\prime j^\prime}  \hat{a}^\dagger_{k} \hat{a}_{m}  \right) }. 
	\end{split}
\end{equation}
The evaluation of matrices $\mathcal{A}$ and $\mathcal{B}$ is more elaborate and it requires a definition of the Hamiltonian; hence we give the details in Appendix \ref{appendix:rpa_commutators}.\

The EOM method is nowadays applied routinely in nuclear physics using excitation operators at the lowest level of complexity. This leads to the so called RPA framework. The RPA, neglecting the coupling to complex internal degrees of freedom, cannot describe collective excitations at a sufficient resolution. An accurate description of collective excitation requires us to consider collective operators that includes higher order multi-body effects. The simplest  straightforward extension of the RPA is the second RPA (SRPA)~\cite{gambacurta2016second, cao2009effects, drozdz1990nuclear}. However, even at the second order, the application of the generalized EOM is computationally demanding on a classical computer due to the increase of the Hilbert space. Thus, the Quantum Equation of Motion (qEOM) seeks to reduce some of the computational burden from a classic computer, which can be performed efficiently on a quantum computer. This is achieved by:
\begin{enumerate}
    \item Computation of the ground state $\ket{gs}$ using the Variational Quantum Eigensolver (VQE) ~\cite{peruzzo2014variational}. This is a hybrid classical-quantum algorithm that a) uses a parameterized quantum circuit to represent the wavefunction $\ket{\psi(\theta)}$, b) uses a quantum computer to efficiently approximate the expectation value $ E = \braket{\psi(\theta)| \hat{H} |\psi(\theta)}$, and c) uses a classical computer to optimize the set of $\theta$ parameters to minimize the cost function $E$. These steps are done recursively between the quantum computer and classical computer until convergence. 
    \item Once the approximate ground state is obtained using the VQE, we then use it to efficiently compute the commutator expectation values of the matrices $\mathcal{A}, \mathcal{B}, \mathcal{C}, \text{and} \; \mathcal{D}$ on a quantum computer.
    \item Finally, we solve the GEE given by Eq. (\ref{eq_gev}) on a classical computer. Note that for relatively large nuclear systems ($N\gg 1$) and high configuration complexity ($\alpha\gg 1$), solving the GEE could become as difficult as finding the direct diagonalization of the many-body Hamiltonian. A possible way around this hurdle is discussed in Section \ref{sec:summary}. 
\end{enumerate}

More details on the qEOM are given in Refs~\cite{Ollitrault_2020, rizzo2022one}. The traditional approach to solve Eq. (\ref{eq_gev}) in the RPA framework is to approximate the correlated many-body ground state $\ket{gs}$ by employing the \emph{Quasi-Boson Approximation} (QBA), such that the expectation value of an operator $\hat{Q}$ is computed with respect to the uncorrelated HF ground state as 
\begin{equation}\label{eq_RPA_HF}
    \braket{\hat{Q}} = \braket{RPA|\hat{Q}|RPA} \approx \braket{HF|\hat{Q}|HF}.
\end{equation}
However, in the qEOM approach, the correlated RPA ground state is approximated by a parameterized quantum circuit that minimizes the Hamiltonian. VQE is the minimization procedure of this circuit which produces a correlated ground state, such that 
\begin{equation}\label{eq_RPA_VQE}
       \braket{\hat{Q}} = \braket{RPA|\hat{Q}|RPA} \approx \braket{VQE|\hat{Q}|VQE}.
\end{equation}
The same is valid for SRPA and higher-order extensions.
In principle, the ground state computed using VQE is more accurate than that obtained from the QBA, because it includes correlations beyond the HF approximation. Therefore, we expect the results for the ground state from VQE to be more accurate than the classical HF and (S)RPA solutions, at least for systems with a small number of particles. 

\subsection{\label{sec:lipkin} Lipkin-Meshkov-Glick Model}

In 1964 Lipkin, Meshkov, and Glick (LMG) proposed a toy model to serve as a test-bed for approximation techniques for solving the quantum many-body problem~\cite{lipkin1965validity, meshkov1965validity, glick1965validity}. A similar Hamiltonian was used by Fallieros in his Ph.D. dissertation in 1959~\cite{fallieros1959collective}. According to LMG, the goal is to have a model that is simple enough to have an exact solution for some cases but also includes non-trivial many-body interactions.  The model has since been used as one of the standard benchmarks for many-body methods in nuclear, condensed matter, and particle physics. Some of the many-body methods tested on this model include the mean-field theory, the random phase approximation~\cite{bohm1953collective}, and the Bardeen-Cooper-Schrieffer theory~\cite{bardeen1957theory, holzwarth1973four}. \

The LMG model describes a system of $N$ interacting fermions constrained on two levels with energies $E = \pm\epsilon/2$. Each energy level is $N$-fold degenerate and the particles interact via a monopole-monopole force. In the quasi-spin formulation, the Hamiltonian is given by
\begin{equation}\label{eq_LMG}
	\hat{H} = \epsilon \hat{J}_z -\frac{V}{2} \left( \hat{J}_+^2 + \hat{J}_{-}^2\right) -\frac{W}{2} \left( \hat{J}_+\hat{J}_{-} + \hat{J}_{-}\hat{J}_+\right),
\end{equation}
where the operators $\hat{J}_z$ and $\hat{J}_{\pm}$ satisfy the angular momentum commutation relations. The interaction term associated with $V$ scatters two particles from the same level up or down, and similarly $W$ simultaneously scatters one particle up and another down or vice versa from different energy levels. The symmetries of this model can be exploited to significantly reduce the size of the relevant Hilbert space. To get a sense of the extent this Hilbert space may be reduced, we compare Eq. (\ref{eq_LMG}) with a general many-body Hamiltonian with up to two-body interaction terms given by
\begin{equation}\label{eq_genHam}
    \hat{H} = \sum_{ij} t_{ij}\hat{a}^\dagger_i \hat{a}_j + \frac{1}{4}\sum_{mnij} \Bar{v}_{mnij} \hat{a}^\dagger_m \hat{a}^\dagger_n \hat{a}_j \hat{a}_i. 
\end{equation}
The full Fock space has dimension $2^N \times 2^N$. On a classical computer, the reduced space with $\Omega$ particles has the dimension $C^{\Omega}_N \times C^{\Omega}_N$. One can further reduce the complexity by noting that the problem is invariant under the exchange of particles within the set of two levels. This is the essence of the \textbf{I} encoding scheme described in Sec. (\ref{sec:I_scheme}). By setting $W=0$ in Eq. (\ref{eq_LMG}), we further realize another symmetry, namely that the interaction term only couples states that differ by spin $M\pm 2$, hence we can block-diagonalize the Hamiltonian.  This leads to the efficient \textbf{J} encoding scheme described in Sec. (\ref{sec:efficient}). In the following, we consider $\Omega=N$, then the problem reduces to the diagonalization of smaller matrices of dimensions $(2J+1) \times (2J+1)$,  where $J=\frac{1}{2}N$. Therefore, the LMG model has an $\mathcal{O}(N)$ complexity which is manageable for classical computers. This is what is needed for a test-bed to benchmark the accuracy of quantum algorithms. However, we must bear in mind that we seek for quantum algorithms that are in principle scalable to be able to solve the general many-body problem with Hamiltonians like Eq. (\ref{eq_genHam}) having arbitrary forms of interactions. It is still unclear whether the qEOM at its current form satisfies this desideratum. \

To get the exact analytical solution of the LMG model for small N values, we consider the eigenstates $\ket{J,M}$ of the operators $\hat{J}_z$ and $\hat{J}^2 = \frac{1}{2} \lbrace \hat{J}_{+}, \hat{J}_{-} \rbrace + \hat{J}_z^2$ as a basis. The quantum numbers are 
${\bf J} = {\bf j}_1 + {\bf j}_2 + \ldots + {\bf j}_N$, which is the total spin, and its projection $M$ in the $z$-direction. The Schr\"{o}dinger's equation 
\begin{equation}\label{schrodinger_eq}
	\hat{H} \ket{\psi} = E \ket{\psi}  
\end{equation}
can be solved with LMG Hamiltonian given in Eq. (\ref{eq_LMG}) using the basis where
\begin{equation}
	\ket{\psi} = \sum^{J}_{M=-J} C_M \ket{J,M}.
\end{equation}
Multiplying Eq. (\ref{schrodinger_eq}) by $\bra{J, M^\prime}$ leads to
\begin{equation}
	\sum_M C_M \bra{J,M^\prime} \hat{H} \ket{J,M} =  C_{M^\prime} E_{JM^\prime},
\end{equation}
and the non-zero matrix elements of $\bra{J,M^\prime} \hat{H} \ket{J,M}$ are given by
\begin{equation}
	\bra{J,M} \hat{H} \ket{J,M} = \epsilon M - W\left[  J(J+1) -M^2\right],
\end{equation}
\begin{equation}
	\begin{split}
			\bra{J,M} \hat{H} \ket{J,M \pm 2} &= \bra{J,M \pm 2} \hat{H} \ket{J,M} \\
			&= -\frac{1}{2} V\times F_{\pm}.
	\end{split}
	\label{MEs}
\end{equation}
The factors $F_{\pm }$ in Eq. (\ref{MEs}) read:
\begin{equation}\label{eq_J_Factor}
	\begin{split}
	F_{\pm } &=\lbrace  \left[J(J+1) -M(M\pm 1)\right] \\
		&\times \left[J(J+1) - (M\pm1)(M\pm2)\right] \rbrace^{\frac{1}{2}}.
	\end{split}
\end{equation}
For a system of $N=2$ particles, the maximum $J=\frac{1}{2}N=1$ and $M=\lbrace -1,0,1 \rbrace$. The Hamiltonian has the dimension $D =3$ and is given by~\cite{co2015hartree}
\begin{equation}
	\hat{H}^{(2)} = 
 \begin{pmatrix}
	\epsilon - W & 0 & -V \\
	0 & -2W & 0 \\
	-V & 0 & -(\epsilon+W) 
\end{pmatrix}.
\end{equation}
This matrix can be diagonalized to get the energy eigenvalues and associated eigenvectors to be
\begin{equation}\label{eq_exact}
	E^{(2)}, \quad \ket{J,M}  =
	\begin{cases}
		+\sqrt{\epsilon^2 + V^2} - W, \qquad  \ket{1,1} \\
		-2W,         \; \;\;    \qquad  \qquad  \qquad \ket{1,0} \\
		-\sqrt{\epsilon^2 + V^2} - W, \qquad  \ket{1,-1}
	\end{cases},    
\end{equation}
which corresponds to both particles in the upper level, one in upper and one in lower level, and both in lower level, respectively. The exact analytical solution for systems with $N>2$ particles is given in Refs.~\cite{lipkin1965validity, co2015hartree}. Some extensions of the LMG model have been proposed, such as the Agassi model~\cite{agassi1968validity, perez2022digital} and the generalized Lipkin model~\cite{carrasco2016generalized}. \

For comparison, we compute the Hartree-Fock solution of the LMG ground state energy given by~\cite{Hagino2000, co2015hartree}
\begin{equation}
E_{\mathrm{HF}}=-\frac{N}{2}\left\{\begin{array}{ll}
\epsilon & \text{for} \,\,\, \tilde{v} < 1 \\
\frac{\epsilon^{2}+(N-1)^{2}V^{2}}{2(N-1)V} & \text{for} \,\,\,  \tilde{v}> 1
\end{array}\right\},
\end{equation}
where $\tilde{v} = V(N-1)/\epsilon$ is the effective interaction strength. Similarly, the RPA solution for the LMG ground state energy is given by~\cite{Hagino2000, co2015hartree}
\begin{equation}
    E_{\mathrm{RPA}} = E_{\mathrm{HF}} + \frac{\omega - A}{2},
\end{equation}
where $\omega = \sqrt{A^2 -|B|^2}$, and $A$ and $B$ read:
\begin{equation}
A =\left\{\begin{array}{ll}
\epsilon & \text{for} \,\,\, \tilde{v} < 1 \\
\frac{3(N-1)^{2}V^{2}}{2(N-1)V - \epsilon^{2}} & \text{for} \,\,\, \tilde{v} > 1
\end{array}\right. ,
\end{equation}
\begin{equation}
B =\left\{\begin{array}{ll}
-(N-1)V & \text{for} \,\,\, \tilde{v} < 1 \\
-\frac{(N-1)^{2}V^{2}}{2(N-1)V + \epsilon^{2}} & \text{for} \,\,\, \tilde{v} > 1
\end{array}\right. .
\end{equation}
We note that the HF and RPA solutions have a discontinuity at $\tilde{v} = 1$, hence this value sets the boundary between the weak and strong coupling regions. 

\subsection{\label{sec:encoding} Encoding schemes}
There are multiple ways we can encode the LMG model on a circuit-based digital quantum computer. In this section we will describe two methods associated with different bases and symmetries used to reduce the relevant Hilbert space.

\subsubsection{Occupation number basis}\label{sec:F_scheme}

Since the LMG model describes a two energy level system with $N$-fold degeneracy, we express the states of the system in terms of occupations numbers in Fock space. Thus, the Hamiltonian given by Eq. (\ref{eq_LMG}) can be written in terms of the creation and annihilation operators by applying the following mappings
\begin{equation}\label{eq_JZops}
	\begin{split}
	&\hat{J}_z  = \frac{1}{2} \sum_{p=1}^N \left( \hat{a}^\dagger_{p,+} \hat{a}_{p,+} - \hat{a}^\dagger_{p,-} \hat{a}_{p,-} \right),  \\
	\end{split}
\end{equation}

\begin{equation}\label{eq_Jops}
	\begin{split}
&\hat{J}_{+} = \sum_{p=1}^N  \hat{a}^\dagger_{p,+} \hat{a}_{p,-} \, \, \text{and} \, \,  \hat{J}_{-} = \left( 	\hat{J}_{+}   \right)^\dagger,
	\end{split}
\end{equation}
where the summation label $p$ represents the set of quantum numbers defining a single-particle state in each energy level. Using Eq. (\ref{eq_JZops}) we can explicitly write the one-body term as
\begin{equation}\label{eq_H0}
	\hat{H}_0 = \epsilon \hat{J}_z = \frac{\epsilon}{2} \sum_{\phi=\pm 1}\sum_{p=1}^N \phi \hat{a}^\dagger_{p,\phi} \hat{a}_{p,\phi},
\end{equation}
where $\phi=\pm 1$ labels the upper and lower energy levels. We then seek to combine indices $p$ and $\phi$ into one index. The binary values $\lbrace -1, +1 \rbrace$ for $\phi$ can be replaced by $\lbrace 0,1 \rbrace$, and the range of values for $p$ can be shifted to $[0,N-1]$. Substituting these changes of the indices into Eq. ($\ref{eq_H0}$), we obtain
\begin{equation}\label{eq_H0s}
	\begin{split}
			\hat{H}_0 &=  \frac{\epsilon}{2} \sum_{\phi=0}^{1}\sum_{p=0}^{N-1} (-1)^ \phi \hat{a}^\dagger_{p,\phi} \hat{a}_{p,\phi}. \\
	\end{split}
\end{equation}

It is easy to see that $p$ and $\phi$ can now be combined into one index with values $[0, 2N-1]$. For clarity, we separate the summation over the hole ($s^\prime$) states with values $[0,N-1]$ and particle ($s$) states with values $[N, 2N-1]$. Therefore, we can rewrite Eq. (\ref{eq_H0s}) as 
\begin{equation}
	\hat{H}_0 = \frac{\epsilon}{2} \left(  \sum_{s^\prime=0}^{N-1} \hat{a}^\dagger_{s^\prime} \hat{a}_{s^\prime}  - \sum_{s=N}^{2N-1} \hat{a}^\dagger_{s} \hat{a}_{s}   \right) 
\end{equation}

To map the two-body terms of the Hamiltonian in Eq. (\ref{eq_LMG}) we use Eq. (\ref{eq_Jops}) to evaluate the products of quasispin operators as follows:
\begin{equation}\label{eq_JsqOp}
		\begin{split}
		&\hat{J}^2_{+} = \sum_{p_1,p_2}  \hat{a}^\dagger_{p_1,+} \hat{a}^\dagger_{p_2,+} \hat{a}_{p_2,-} \hat{a}_{p_1,-}  \\
		&\hat{J}_{+}\hat{J}_{-}  =  \sum_{p_1,p_2}  \hat{a}^\dagger_{p_1,+} \hat{a}^\dagger_{p_2,-} \hat{a}_{p_2,+} \hat{a}_{p_1,-} . \\
	\end{split}
\end{equation}
Hence, we can write the $V$-scattering term as 
\begin{equation}\label{eq_Hv}
	\begin{split}
		\hat{H}_v &=  -\frac{V}{2} \left( \hat{J}_+^2 + \hat{J}_{-}^2\right) \\
		&= -\frac{V}{2}  \sum_{q,r} \sum_{q^\prime,r^\prime} \left( \hat{a}^\dagger_{q} \hat{a}^\dagger_{r} \hat{a}_{r^\prime} \hat{a}_{q^\prime}  + \hat{a}^\dagger_{q^\prime} \hat{a}^\dagger_{r^\prime} \hat{a}_{r} \hat{a}_{q}  \right) \Delta^{q,q^\prime}_{r,r\prime},	
	\end{split}
\end{equation}

and the $W$-scattering one as
\begin{equation}\label{eq_Hw}
	\begin{split}
		\hat{H}_w &=  -\frac{W}{2} \left( \hat{J}_+\hat{J}_{-} + \hat{J}_{-}\hat{J}_+\right) \\
		&= -\frac{W}{2}  \sum_{q,r} \sum_{q^\prime,r^\prime} \left( \hat{a}^\dagger_{q} \hat{a}^\dagger_{r^\prime} \hat{a}_{r} \hat{a}_{q^\prime}  + \hat{a}^\dagger_{q^\prime} \hat{a}^\dagger_{r} \hat{a}_{r^\prime} \hat{a}_{q}  \right) \Delta^{q,q^\prime}_{r,r\prime}, 	
	\end{split}
\end{equation}
where the constraint $ \Delta^{q,q^\prime}_{r,r\prime} = \delta_{q+N,q^\prime } \delta_{r+N,r^\prime }$ is added to keep the symmetry of the LMG model. The new final Hamiltonian is the sum of all the terms given above
\begin{equation} \label{eq_lmg_FS}
	\hat{H}_F = \hat{H}_0 + \hat{H}_v + \hat{H}_w,
\end{equation}
where the subscript $F$ emphasizes that it is defined in the Fock-space. To encode this Hamiltonian on a quantum computer, we can use the Jordan-Wigner transformation~\cite{jordan1993paulische, nielsen2005fermionic} to convert the fermionic operators into qubit operators such that
\begin{equation}\label{eq_FtoQ}
	\hat{H}_{F} \left( \hat{a}^\dagger, \hat{a} \right) \rightarrow \hat{H}_{Q}\left(\sigma^{\pm},\sigma^i\right), 
\end{equation}
where $\sigma^i$ defines the Pauli matrices for $i=\lbrace 0,1,2,3\rbrace \rightarrow \lbrace \mathbb{I}, X,Y,Z \rbrace$ and $\sigma^{\pm} = X \pm iY$. The corresponding basis states for the two-level LMG model after this mapping are given by
\begin{equation}\label{eq_st_map}
	\ket{\eta_{N-} \ldots \eta_{1-}, \eta_{N+} \ldots \eta_{1+}} \longrightarrow \ket{q_{2N-1} \ldots q_0},
\end{equation}
with $\eta \in [0,1]$, which represents an empty or occupied fermionic state, and $q \in [0,1] $ represents a spin up or down qubit state. One of the states of a system with $N=2$ particles can be explicitly written as
\begin{equation}\label{eq_state_F}
	\ket{1_{-} 1_{-}, 0_{+} 0_{+}} \longrightarrow \ket{1100},
\end{equation}
which corresponds to the uncorrelated ground state.  In this scheme, the Hamiltonian is encoded in the full Fock-space that has a size of $2^{2N}$. Thus, many states are not used leading to a large dark sector (unused space). In the ensuing subsections we discuss more efficient encoding schemes that exploit symmetries of the LMG model.

\subsubsection{Individual spin basis}\label{sec:I_scheme}

Since the LMG Hamiltonian is invariant under the exchange of particles within the set of two levels, we can exploit this symmetry to 
reduce the number of states by a factor of two.
This can be naturally seen by considering the basis of the individual spin $\ket{{\bf j}_1, {\bf j}_2, \ldots , {\bf j}_N}$ of the particles, where ${\bf j} = \pm \frac{1}{2}$. This can be straightforwardly mapped to qubit basis. For $N=2$, the spin eigenstates are mapped to qubits as follows
\begin{equation}\label{eq_state_I}
	\ket{\uparrow \uparrow}, \ket{\uparrow \downarrow}, \ket{\downarrow \uparrow} \ket{\downarrow \downarrow} \longrightarrow \ket{11}, \ket{10}, \ket{01}, \ket{00}.
\end{equation}

To transform the Hamiltonian given by Eq. (\ref{eq_LMG}) into linear products of Pauli matrices, we simply express it in the individual spin basis by applying the following conversion ~\cite{cervia2021lipkin}
\begin{equation}\label{eq_jZops}
	\begin{split}
	&\hat{J}_z  = \frac{1}{2} \sum_{p}  \, \hat{j}^{(p)}_z ,  \\
	&\hat{j}^{(p)}_z = \hat{a}^\dagger_{p,+} \hat{a}_{p,+} - \hat{a}^\dagger_{p,-} \hat{a}_{p,-}  \; ,
	\end{split}
\end{equation}
for the non-interacting term and, similarly,
\begin{equation}\label{eq_jops}
	\begin{split}
&\hat{J}_{+} = \sum_{p} \; \hat{j}^{(p)}_{+}  \, \, \text{and} \, \,  \hat{J}_{-} = \left( 	\hat{J}_{+}   \right)^\dagger \\
&\hat{j}^{(p)}_{+} = \hat{a}^\dagger_{p,+} \hat{a}_{p, -} \, \, \text{and} \, \, \hat{j}^{(p)}_{-} = \hat{a}^\dagger_{p,-} \hat{a}_{p,+}
	\end{split},
\end{equation}
for the interacting terms. We substitute the operators from Eqs. (\ref{eq_jZops}, \ref{eq_jops}) into the Hamiltonian given by Eq. (\ref{eq_LMG}) to get
\begin{equation}\label{eq_LMG_I}
\begin{split}
    \hat{H}_{I} &= \frac{\epsilon}{2}  \sum_{p}  \, \hat{j}^{(p)}_z - \frac{V}{2} \sum_{p \neq q} \left( \hat{j}^{(p)}_{+}\hat{j}^{(q)}_{+} + \hat{j}^{(q)}_{-}\hat{j}^{(p)}_{-} \right)\\
    & - \frac{W}{2} \sum_{p \neq q} \left( \hat{j}^{(p)}_{+}\hat{j}^{(q)}_{-} + \hat{j}^{(p)}_{-}\hat{j}^{(q)}_{+}\right)
\end{split}.
\end{equation}
Note that in this representation the Hamiltonian $\hat{H}_{I}$ has a Hilbert space size of  $2^{N}$. From this point on-wards we use the dimensionless Hamiltonian $\Bar{H} = \hat{H}/\epsilon$ with $W=0$ and the interaction strength given by
\begin{equation}
    v = V/\epsilon .
\end{equation}

As an example, we consider a system of $N=2$ particles where the Hamiltonian can be explicitly written as
\begin{equation} \label{eq_LMG_I_N2}
\begin{split}
    \Bar{H}^{(2)}_I &= \frac{1}{2}  \left( \hat{j}^{(1)}_z + \hat{j}^{(2)}_z \right) - \frac{v}{2} \left(  \hat{j}^{(1)}_{+}\hat{j}^{(2)}_{+} + \hat{j}^{(2)}_{-}\hat{j}^{(1)}_{-} \right)\\
    &= \frac{1}{2} \left( Z_1  +  Z_2 \right) - \frac{v}{2} \left( X_1 \otimes X_2 - Y_1 \otimes Y_2 \right),
\end{split}
\end{equation}
where $j_z = Z$ and $j_{\pm} = \left( X \pm iY \right)/\sqrt{2}$. The encoding and Hamiltonian form  of Eq. (\ref{eq_LMG_I_N2}) corresponds to Eq. (7) of Ref.~\cite{cervia2021lipkin}. For $N=3$ the Hamiltonian is given by~\cite{cervia2021lipkin}
\begin{equation} \label{eq_LMG_I_N3}
	\begin{split}
		\Bar{H}^{(3)}_I &= \frac{1}{2} \left( Z_1 + Z_2 + Z_3 \right) \\
		&- \frac{v}{2} \left( X_1  X_2 + X_1 X_3 +   X_2  X_3 \right)\\
		&+ \frac{v}{2} \left( Y_1  Y_2 + Y_1 Y_3 +   Y_2  Y_3 \right).
	\end{split}
\end{equation}
Therefore, a LMG system of $N$ particles can be encoded using $N$ qubits in the individual spin basis which is much better than the occupation number basis  requiring $2N$ qubits. It is worth mentioning that this reduction of the number of required qubits by a factor of two follows from the symmetry of the two-level LMG model and may be different for a multi-level LMG model. We can further improve the encoding of the LMG model on a quantum computer by exploiting another symmetry of the LMG Hamiltonian when $W=0$. This leads to a more efficient encoding scheme, described in the ensuing section.

\section{\label{sec:efficient}Efficient Encoding Scheme}

We consider the coupled $\ket{J,M}$ basis used in Sec. (\ref{sec:lipkin}), where $J=\frac{N}{2}$ and $M \in \left[ -J, -J + 1, \ldots , 0, \ldots J-1, J \right]$, thus, the full basis is of size $D = 2J + 1$. We note that by setting $W=0$ in Eq. (\ref{eq_LMG}), another symmetry arises from the interaction term which only couples states that differ by spin $M\pm 2$.  Hence, the Hamiltonian can be 
block-diagonalized, which reduces the number of the "relevant states" to at most $d = J + 1 $. These states can be mapped to qubits as follows
\begin{equation}\label{eq_SB_map}
\begin{aligned}
|J,-J\rangle & \equiv|0\rangle \rightarrow \ket{\text{bin}(0)}, \\
|J,-J+2\rangle & \equiv|1\rangle \rightarrow \ket{\text{bin}(1)},\\
\cdots & \\
|J,J-2\rangle & \equiv\left|d-2\right\rangle \rightarrow \ket{\text{bin}(d-2)}, \\
|J,J\rangle & \equiv\left|d-1\right\rangle \rightarrow \ket{\text{bin}(d-1)},
\end{aligned}
\end{equation}
where $\ket{\text{bin}(k)} \equiv \ket{q_1,q_2, \ldots , q_n}$,  $k = \sum_{i=1}^n q_i 2^{n-i}$ with $q_i \in \{0,1\}$. This mapping method is sometimes called the Standard Binary (SB) encoding. For Hamiltonian simulations on a quantum computer, a more efficient encoding than the SB is the Gray code (GC) ~\cite{sawaya2020resource,di2021improving}, which effectively uses less gates and a lower circuit depth. The Gray code is defined to be an ordering of binary values where any two adjacent entries differ by only a single bit \cite{gray1953pulse}. For example, consider the eight states of three binary bits, which can be ordered sequentially as 
\begin{equation}\label{eq_GC_map}
\begin{aligned}
&0 \rightarrow \ket{000}, 1 \rightarrow \ket{00{\bf 1}}, 2 \rightarrow \ket{0{\bf 1}1}, 3 \rightarrow \ket{01{\bf 0}},\\
&4 \rightarrow \ket{{\bf 1}10}, 5 \rightarrow \ket{11{\bf 1}}, 6 \rightarrow \ket{1{\bf 0}1}, 7 \rightarrow \ket{10{\bf 0}},
\end{aligned}   
\end{equation}
where the single bit that changes between adjacent states is shown in bold. A set of Gray code with $\nu$ bits is expressed as
\begin{equation}
    \mathbf{G}_{\nu} = \lbrace g_0, g_1, \ldots g_{2^\nu-1} \rbrace,
\end{equation}
where each $g_{i}$ is a sequence of $\nu$ bits. Thus, with this encoding we can write the Hamiltonian as 
\begin{equation}\label{eq_LMG_J}
    \hat{H}_J = \sum_{k=0}^{d-1} a_k \ket{k}\bra{k} +  \sum_{k=0}^{d-2} b_k \left[ \ket{k}\bra{k+1} + \ket{k+1}\bra{k} \right], 
\end{equation}
with the coefficients
\begin{equation}\label{eq_ak}
    a_k = \epsilon \left[ 2k - J \right] = \epsilon M ,
\end{equation}
\begin{equation}\label{eq_bk}
    b_k = -\frac{V}{2} \times F_{+}(M=2k - J),
\end{equation}
where the function $F_{+}$ is defined in Eq. (\ref{eq_J_Factor}). We illustrate below how our encoding scheme works for $N=4$ and then generalize to arbitrary $N$.

\subsection{N=4}\label{sec:example_N4}
As an illustration, we first consider a system of $N=4$ particles where $J=2$, and for even values of $M$ we get three states, which can be encoded as
\begin{equation}\label{eq_GC2_map}
\begin{aligned}
|2,-2\rangle & \equiv|0\rangle \rightarrow \ket{00}, \\
|2,0\rangle & \equiv|1\rangle \rightarrow \ket{01},\\
|2,+2 \rangle & \equiv|2\rangle \rightarrow \ket{11}.
\end{aligned} 
\end{equation}
The associated Hamiltonian is given by 
\begin{equation}\label{eq_LMG_J_4}
\begin{split}
\Bar{H}^{(4)}_{Je} &= a_0 \ket{00}\bra{00} + a_1 \ket{01}\bra{01} + a_2  \ket{11}\bra{11} \\
&+ b_0 \left[ \ket{00}\bra{01} + \ket{01}\bra{00} \right] \\
&+ b_1 \left[ \ket{01}\bra{11} + \ket{11}\bra{01} \right],
\end{split}
\end{equation}
where the subscript $"Je"$ represents the \textbf{J}-scheme with $"e"$ for even $M$ values. We can a priori directly write the matrix form of this Hamiltonian as 
\begin{equation}\label{eq_LMG_J_4_matrix}
    \Bar{H}^{(4)}_{Je} = \begin{pmatrix}
    a_0 & b_0 & 0 & 0 \\
    b_0 & a_1 & 0 & b_1 \\ 
    0   & 0   & 0 & 0 \\
    0 & b_1   & 0 & a_2
    \end{pmatrix}_{GC}.
\end{equation}
For comparison the Hamiltonian in the SB basis is given by
\begin{equation}\label{eq_LMG_J_4_matrix_SB}
    \Bar{H}^{(4)}_{Je}  = \begin{pmatrix}
    a_0 & b_0 & 0   & 0 \\
    b_0 & a_1 & b_1 & 0 \\ 
    0   & b_1 & a_2 & 0 \\
    0   & 0   & 0   & 0
    \end{pmatrix}_{SB}.
\end{equation}
Note that the unused state $\ket{10}_{GC}$ or $\ket{11}_{SB}$ is not coupled to the others, thus giving a row and column of zeros on Hamiltonian. This problem arises from the fact that the set of available states on a quantum computer come in powers of two whilst the number of states we wish to encode can be any positive integer. For larger systems, this may introduce spurious solutions.\

We can transform the Hamiltonian of Eq. (\ref{eq_LMG_J_4}) in terms of Pauli matrices by noting that the operators associated with $a_k$ are given by
\begin{equation}\label{eq_ak_4}
\begin{aligned}
\ket{00}\bra{00} &= P^{(0)}_1 P^{(0)}_0 = \frac{1}{4}\left( \mathbb{I} + Z_0 + Z_1 + Z_1 Z_0 \right) \, ,  \\
\ket{01}\bra{01} &= P^{(0)}_1 P^{(1)}_0 = \frac{1}{4}\left( \mathbb{I} - Z_0 + Z_1 - Z_1 Z_0 \right) \, ,  \\
\ket{11}\bra{11} &= P^{(1)}_1 P^{(1)}_0 = \frac{1}{4}\left( \mathbb{I} - Z_0 - Z_1 + Z_1 Z_0 \right) \, ,   
\end{aligned} 
\end{equation}
where $P^{(0)}_i=\frac{1}{2}\left( \mathbb{I}_i + Z_i \right)$ and $P^{(1)}_i=\frac{1}{2}\left( \mathbb{I}_i - Z_i \right)$ are the projection operators acting on the $i^{\text{th}}$ qubit. The operators associated with $b_k$ can be converted to \begin{equation}\label{eq_bk_4}
\begin{aligned}
\ket{00}\bra{01}+ \ket{01}\bra{00}  &= P^{(0)}_1 X_0 = \frac{1}{2}\left( X_0 + Z_1 X_0 \right) \, , \\
\ket{01}\bra{11}+ \ket{11}\bra{01}  &= X_1 P^{(1)}_0 = \frac{1}{2}\left( X_1 - X_1 Z_0 \right) \, .  \\
\end{aligned}
\end{equation}
Note that the order of operations is important. For instance, the gate $Z_0$ should be interpreted as $\mathbb{I}_1 Z_0$ while $Z_1$ is $Z_1 \mathbb{I}_0$, otherwise we do not get the proper matrix form when performing the tensor product. Substituting  Eqs. (\ref{eq_ak_4}) and (\ref{eq_bk_4}) into Eq. (\ref{eq_LMG_J_4}), we get 
\begin{equation}\label{eq_LMG_J_4b}
\begin{split}
\Bar{H}^{(4)}_{Je} &= \frac{1}{4}\left( a_0 + a_1 + a_2 \right)\mathbb{I} + \frac{1}{4}\left( a_0 - a_1 - a_2 \right) Z_0 \\
&+ \frac{1}{4}\left( a_0 + a_1 - a_2 \right) Z_1 + \frac{1}{4}\left( a_0 - a_1 + a_2 \right) Z_1 Z_0 \\
&+ \frac{1}{2}b_0 \left( X_0 + Z_1 X_0 \right) + \frac{1}{2}b_1 \left( X_1 - X_1 Z_0 \right).
\end{split}    
\end{equation}
Using Eqs. (\ref{eq_ak}) and (\ref{eq_bk}), we find the Hamiltonian coefficients to be  
\begin{equation}
\begin{aligned}
 &a_0 = -2 \; , a_1 = 0 \; , a_2 = 2  \; , \\
 &b_0 = b_1 = - v\sqrt{6} .
\end{aligned}
\end{equation}
Therefore, Eq. (\ref{eq_LMG_J_4b}) can be written as 
\begin{equation}\label{eq_LMG_J_N4_Meven}
\begin{split}
\Bar{H}^{(4)}_{Je} &=  -\left( Z_0 + Z_1 \right)
-\frac{\sqrt{6}}{2} v \left( X_0 + X_1 + Z_1 X_0 - X_1 Z_0\right)\\
&=\begin{pmatrix}
    -2 & -\sqrt{6}v & 0 & 0 \\
    -\sqrt{6}v & 0  & 0 & -\sqrt{6}v \\ 
    0   & 0    & 0 & 0 \\
    0 & -\sqrt{6}v    & 0 & -2
    \end{pmatrix}.
\end{split}    
\end{equation}
By diagonalizing Eq. (\ref{eq_LMG_J_N4_Meven}), we obtain the energy solutions $\Bar{E}^{(4)} = \lbrace 0, \; \pm 2\sqrt{3v^2 + 1} \rbrace$. Comparing with the exact analytical solution of Ref. \cite{co2015hartree}, where the energy spectrum for $N=4$ is given by
\begin{equation}
 \Bar{E}^{(4)} = \lbrace 0, \; \pm \sqrt{9v^2 + 1}, \; \pm 2\sqrt{3v^2 + 1} \rbrace,
\end{equation}
we note that we are missing two solutions. These remaining solutions are found by considering the two states with odd values of $M$, which can be mapped onto one qubit as follows:
\begin{equation}\label{eq_N4_map2}
\begin{aligned}
\ket{2, -1} & \equiv \ket{0}, \qquad
\ket{2, +1} & \equiv \ket{1}
\end{aligned}. 
\end{equation}
The associated Hamiltonian for the odd values of $M$ can be constructed similarly to the even $M$, where $a_1 = -a_0 = 1$ and $b_0 = -3v$, to get 
\begin{equation}\label{eq_LMG_J_N4_Modd}
\begin{split}
    \Bar{H}^{(4)}_{Jo} &= -Z -3v X = \begin{pmatrix}
    -1 & -3v \\
    -3v & 1 
    \end{pmatrix},
\end{split}
\end{equation}
where the $"Jo"$ in the subscript stands for odd values of $M$ in the \textbf{J}-scheme. It is straightforward to see that diagonalizing Eq. (\ref{eq_LMG_J_N4_Modd}) gives us the two remaining energy solutions $\Bar{E}^{(4)} = \pm \sqrt{9v^2 + 1}$. \

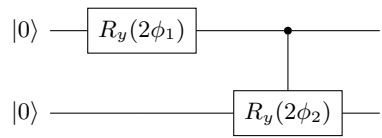
\begin{figure}[h!]
\centering
    \begin{quantikz}[thin lines]
    \lstick{$\ket{0}$} & \gate{R_y(2\phi_{1})}  & \ctrl{1} & \qw \\
    \lstick{$\ket{0}$} & \qw & \gate{R_y(2\phi_{2})} & \qw 
    \end{quantikz}
\caption{N=4, J-scheme ansatz for even M values.}
\label{fig:J_N4_ansatz_A}
\end{figure}

To construct the associated quantum circuit for the system with $N=4$, we look at its wave function which can be split into two sets 
\begin{equation}
\ket{\psi_J}=
\begin{cases}
\ket{\psi_e} = c_{0e}\ket{2,2} + c_{1e} \ket{2,0} + c_{2e} \ket{2,-2}, \\
\ket{\psi_o} = c_{0o}\ket{2,-1} + c_{1o} \ket{2,+1}. 
\end{cases}
\end{equation}
Using the Gray encoding, the wave function for even $M$ values is given by 
\begin{equation}
\begin{split}
\ket{\psi_e(\phi_1, \phi_2)} &=\cos{\phi_1}\ket{00} + \cos{\phi_2}\sin{\phi_1}\ket{01} \\
&+ \sin{\phi_2}\sin{\phi_1}\ket{11}, 
\end{split}
\end{equation}
which is represented by the quantum circuit shown in Fig. \ref{fig:J_N4_ansatz_A} where the gate $R_y(\phi) = \exp{\left(-i\frac{\phi}{2} Y \right)}$ and $\phi \in [0, \frac{\pi}{2})$.

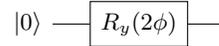
\begin{figure}[h!]
\centering
    \begin{quantikz}[thin lines]
    \lstick{$\ket{0}$} & \gate{R_y(2\phi)} & \qw 
    \end{quantikz}
\caption{N=4, J-scheme ansatz for odd M values.}
\label{fig:J_N4_ansatz_B}
\end{figure}

Similarly, the wave function for odd $M$ values is given by 
\begin{equation}
\ket{\psi_o(\phi)} =\cos{\phi}\ket{0} + \sin{\phi}\ket{1},
\end{equation}
which is represented by the quantum circuit shown in Fig. \ref{fig:J_N4_ansatz_B}.

For comparison, we consider the ground state wave function for the \textbf{I}-scheme~\cite{cervia2021lipkin}, which is a superposition of eight states given by
\begin{equation}
\begin{aligned}
|\psi_I(\theta)\rangle &= \cos ^{2} \theta|\downarrow \downarrow \downarrow \downarrow\rangle+\sin ^{2} \theta|\uparrow \uparrow \uparrow \uparrow\rangle \\
&-\frac{1}{\sqrt{12}} \sin 2 \theta(|\uparrow \uparrow \downarrow \downarrow\rangle+|\downarrow \downarrow \uparrow \uparrow\rangle+|\downarrow \uparrow \downarrow \uparrow\rangle \\
&+|\downarrow \uparrow \uparrow \downarrow\rangle+|\uparrow \downarrow \downarrow \uparrow\rangle+|\uparrow \downarrow \uparrow \downarrow\rangle),
\end{aligned}
\end{equation}
where $\theta \in [0, \frac{\pi}{2})$. The associated quantum circuit would require four qubits and at least seven gates. Therefore, our encoding scheme uses much less quantum resources than the \textbf{I}-scheme, which becomes much more critical for systems with large number of particles. 

\subsection{Arbitrary N}
We note that for the case of $N=4$ we essentially split the Hamiltonian into two decoupled parts, which we diagonalize independently to obtain the complete spectrum. This procedure can be generalized for the case of an arbitrary $N$, where the Hamiltonian is split into a block form as 
\begin{equation}\label{eq_gen_H}
	\Bar{H}^{(N)}_J  = \begin{pmatrix}
	\Bar{H}_A & 0 \\
	0 & \Bar{H}_B
	\end{pmatrix},
\end{equation}
with the block sizes $d_A = J+1$ and $d_B = J$ for the even values of $N$, and $d_A=d_B = \frac{1}{2}(N+1)$ for the odd values of $N$. We can now compare the size of Hilbert space and the number of qubits required for each of the three different encoding schemes:
\begin{equation}
\begin{aligned}
 &\textbf{F}-\text{scheme} : d_F =2^{2 N} \rightarrow 2N \text{ qubits} \\
& \textbf{I}-\text{scheme} : d_I =2^{N} \rightarrow N \text{ qubits} \\
&\textbf{J}-\text{scheme} : d_J = \left(\frac{N}{2}+1\right) \rightarrow q_N \text{ qubits},
\end{aligned}
\end{equation}
where $q_N$ in the last case is the first integer that satisfies 
\begin{equation}
    q_N \geq \log{_2\left(\frac{N}{2}+1\right)}.   
\end{equation}
This implies, for instance, that the Lipkin model with $N=100$ particles can be solved with at most $q_N = 7$ qubits using our efficient \textbf{J}-scheme, while it would require 100 and 200 qubits for the \textbf{I}-scheme and \textbf{F}-scheme, respectively.

\section{\label{sec:results} Results}

Using the qEOM, we simulate the LMG model for a system with $N=2,3$, and $4$ particles. We compare the results of our efficient encoding (\textbf{J}-scheme) with the individual spin basis (\textbf{I}-scheme) given in Ref. \cite{cervia2021lipkin}. The results for each scheme are marked with the post-factor labels \textbf{J} and \textbf{I} in the legends of the plots of the energy as a function of the interaction strength. The Hamiltonians and circuit ans\"atze used for $N=2, $ and $3$ are given in Appendix \ref{appendix:encoding_schemes}. Depending on availability, we used the IBM quantum devices: \verb|santiago|, \verb|manila|, and \verb|bogota| which all have 5 qubits and a quantum volume (defined in Ref. ~\cite{cross2019validating}) of 32. \

\begin{figure}[h!]
	\centering
	\includegraphics[width=\columnwidth]{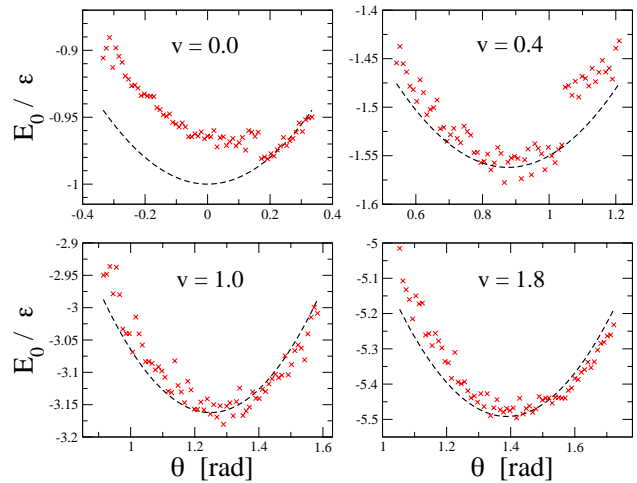}
	\cprotect\caption{Example of line search plot of the ground state energy ($\Bar{E}_0=E_0/\epsilon$) as a function of the wave-function parameter ($\theta$). The scatter (red) "x" points are computed from the quantum device and the dashed (black) line are from a state vector simulator for the $N=4$ J-scheme circuit shown in Fig. \ref{fig:J_N4_ansatz_B}.}
	\label{fig:line_search}
\end{figure}

First, we use the VQE algorithm \cite{peruzzo2014variational} to compute the ground state and its energy for the LMG Hamiltonian. The goal of VQE is to find the optimal set of angles $\lbrace \theta_0 \rbrace$ that minimizes the energy given by
\begin{equation}
	\Bar{E}({\theta}) = \bra{\psi({\theta})} \hat{H} \ket{\psi({\theta})},
\end{equation}
where $\Bar{E} = E/\epsilon$. Usually the optimization of $\lbrace \theta \rbrace$ requires a computation of derivatives of $\Bar{E}({\theta})$ which can be difficult for a large set of parameters. Here, since only a few angles need to be optimized, a more direct approach is used. For circuits with one angle, we do a line search by computing $\Bar{E}({\theta})$ for various angles within the domain of $\theta \in [0, \frac{\pi}{2})$ and then take the minimum energy. We can visualize this method by computing the energy landscape of $\Bar{E}({\theta})$ at various interaction strengths using both a simulator and a quantum computer. As shown in Fig. \ref{fig:line_search}, the results from the quantum computer are fairly close to the simulator ones with errors of $\Delta \theta \leq 0.2$ rad and $\Delta \Bar{E} \leq 0.05 $ for the optimal angle and minimum energy, respectively. These errors illustrate the degree of imperfection of current quantum devices. We follow a similar method for circuits with two angles. 

\begin{algorithm}[H]
 \SetAlgoLined
 \KwData{$N=2, 3, 4$}
 \KwResult{$E_n$, $\ket{\psi_n}$}
 \For{v \text{in} \text{vlist}}{
  $\hat{H}_v \leftarrow$ construct LMG Hamiltonian at $v$ \;
  $\Bar{E}_0^\prime, \lbrace \theta^\prime_0 \rbrace \leftarrow$ VQE($\hat{H}_v$, simulator, optimizer) \;
  search $\leftarrow$  search intervals $\left[ \theta^\prime_0 - \delta,  \; \theta^\prime_0 + \delta  \right]$ \;
  $\Bar{E}_0, \lbrace \theta_0 \rbrace \leftarrow$ VQE($\hat{H}_v$, device, search) \;
  $\ket{\psi_0} \leftarrow$ construct g.s. circuit $U\left( \lbrace \theta_0 \rbrace \right)$ \;
  $\mathcal{A}, \mathcal{B}, \mathcal{C}, \mathcal{D}  \leftarrow$   expectation$\left( \ket{\psi_0}, \hat{K}(\alpha), \hat{H} \right)$  \;
  $E_n, \ket{\psi_n} \leftarrow$ solve GEE$\left( \mathcal{A}, \mathcal{B}, \mathcal{C}, \mathcal{D} \right) $ \; 
 }
 \caption{qEOM for Lipkin model}
\label{algo:qEOM}
\end{algorithm}
Second, we use the qEOM algorithm \cite{Ollitrault_2020} to compute the excited states and energy of the LMG Hamiltonian. We slightly modified some parts of the algorithm to suit our problem as shown in Algorithm \ref{algo:qEOM}. We set $vlist = [0, 2]$ which covers the weak and strong coupling regimes. We use the Limited memory and bounded Broyden–Fletcher–Goldfarb–Shanno  (L-BFGS-B)~\cite{zhu1995limited} optimizer for running VQE on the simulator. The L-BFGS-B is a quasi-Newton method that approximates the Hessian matrix (second-order differentials) based on successive iterations and does not need to store the entire Hessian, which reduces the computer memory required and allows bounds to be set for the variable parameter values. The results are given in the ensuing subsections.
 
\subsection{Ground State Energy}

\begin{figure}[h!]
	\centering
	\includegraphics[width=\columnwidth]{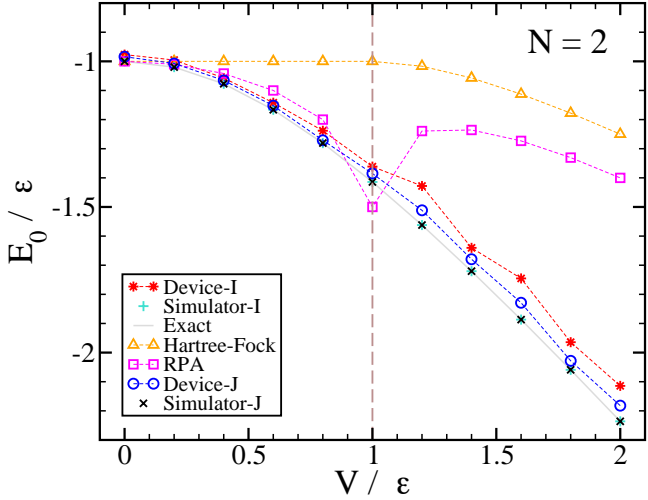}
	\cprotect\caption{Ground state energy ($\Bar{E}_0=E_0 / \epsilon$) as a function of the  interaction strength ($v=V/ \epsilon$) for a LMG system of $N=2$ particles. }
	\label{fig:GS_Energy_N2}
\end{figure}

\begin{figure}[h!]
	\centering
	\includegraphics[width=\columnwidth]{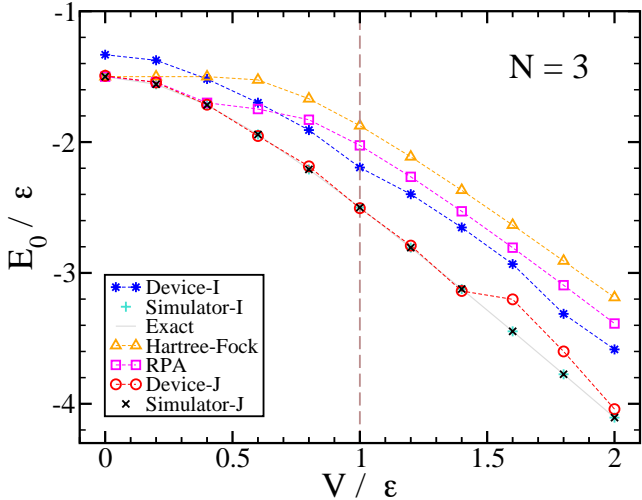}
	\cprotect\caption{Ground state energy ($\Bar{E}_0=E_0/\epsilon$) as a function of the interaction strength ($v=V/\epsilon$) for a LMG system of $N=3$ particles.}
	\label{fig:GS_Energy_N3}
\end{figure}

\begin{figure}[h!]
	\centering
	\includegraphics[width=\columnwidth]{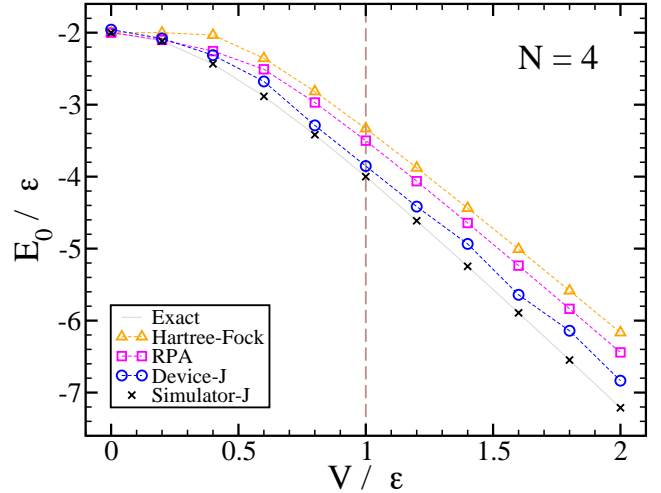}
	\cprotect\caption{Ground state energy ($\Bar{E}_0=E_0/\epsilon$) as a function of the interaction strength ($v=V/\epsilon$) for a LMG system of $N=4$ particles.}
	\label{fig:GS_Energy_N4}
\end{figure}

We compute the ground state energy of the LMG Hamiltonian as a function of the interaction strength in both weak ($v<1)$ and strong $(v\geq 1)$ coupling regimes. Note that we redefine the borderline between weak and strong coupling. 
A comparison of the VQE solution using IBM quantum computer is made with the exact analytical solution, the classical Hartree-Fock, and RPA solution as shown in Figs. \ref{fig:GS_Energy_N2} - \ref{fig:GS_Energy_N4}. We observe that the results deviate from each other in both weak and strong coupling regimes. For all cases, the simulator results are almost identical to the exact solutions, which means that all computational errors can be attributed to noise in the quantum device. No error mitigation method was performed for this work as we were interested in comparing the raw results to be able to investigate the effects of increasing the model parameters $\lbrace N, v, \alpha \rbrace$ for both the \textbf{I}- and \textbf{J}-schemes.  

For N=2, the ground state energy as a function of the interaction strength is shown in Fig. \ref{fig:GS_Energy_N2}. In the weak coupling regime, the VQE results for both the \textbf{I}- and \textbf{J}-schemes are relatively close to the exact solution and RPA solutions with average errors of about $2\%$ and  $1\%$ respectively (see Appendix \ref{appendix:av_errors}). In the strong coupling regime, the results of the \textbf{J}-scheme are slightly more accurate than those of the \textbf{I}-scheme, but both are relatively close to the exact solution. In both regions, the VQE solution from both schemes was significantly more accurate than both HF and RPA. This was expected as the classical HF and RPA perform better for systems with large number of particles. 

For N=3, the ground state energy as a function of the interaction strength is shown in Fig. \ref{fig:GS_Energy_N3}.
In the weak coupling regime, the VQE solution for the \textbf{J}-scheme is relatively close to the exact solution, whilst the \textbf{I}-scheme solution slightly deviates from it. For some values of $v$, namely $v \leq 0.6$ for the RPA and  $v<0.6$ for HF, the VQE solution for the \textbf{I}-scheme is less accurate than the HF and RPA, but much better for $v>0.6$. In the strong coupling regime, in most cases the VQE solution for the \textbf{J}-scheme is relatively close to the exact solution with minor deviations on a few points. However, the \textbf{I}-scheme results significantly deviate from the exact solution with average errors of about $19\%$ but still remains slightly more accurate than both the HF and RPA solutions. For all regions, the \textbf{J}-scheme is more accurate than the \textbf{I}-scheme, HF, and RPA solutions. These results can be understood by noting that, for this simulation, the \textbf{J}-scheme only used one qubit and one single-qubit gate (see Appendix \ref{appendix:encoding_schemes}), that accumulates less errors on a quantum computer than the \textbf{I}-scheme which required three qubits and seven gates including three CNOT-gates. 

For N=4, the ground state energy as a function of the  interaction strength is shown in Fig. \ref{fig:GS_Energy_N4}. The circuit for the \textbf{I}-scheme is relatively complex, thus its simulation was omitted in this work. For both the weak and strong coupling regimes, the VQE solution for the \textbf{J}-scheme is relatively close to the exact solution with small deviation in the strong coupling region. Also the \textbf{J}-scheme is more accurate than both the HF and RPA solutions at all values of the interaction strength.\

We also observe that, for all particle numbers investigated in this work as shown in Figs. \ref{fig:GS_Energy_N2} - \ref{fig:GS_Energy_N4}, the VQE solution for the ground state energy for both encoding schemes is generally more accurate than the classical HF and RPA solutions. The VQE results generally have larger errors in the strong coupling regime, which is consistent with the findings of Ref~\cite{Ollitrault_2020}. Such errors can be combated by employing error mitigation methods and using better quantum computers with high qubit quality.

\subsection{Excited State Energies}

As described in Sec. (\ref{sec:qEOM}), the next step after VQE in the qEOM method is to compute the excited states and their energies by solving Eq. (\ref{eq_gev}). We computed the energy spectrum of the LMG Hamiltonian as a function of the interaction strength for $N=2, 3$, and $4$ particles. The results of the qEOM runs on the IBM quantum computer were compared with the exact analytical solutions for both the \textbf{I}- and \textbf{J}-schemes. 

\begin{figure}[h!]
	\centering
	\includegraphics[width=\columnwidth]{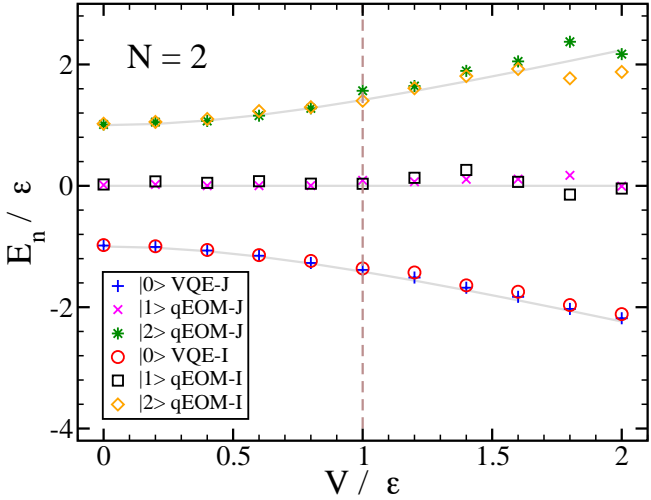}
	\cprotect\caption{Energy spectrum ($ \Bar{E}_n=E_n/\epsilon$) as a function of the interaction strength ($v=V/\epsilon$) for $N=2$ particles.}
	\label{fig:Energy_Spectrum_N2}
\end{figure}

\begin{figure}[h]
	\centering
	\includegraphics[width=\columnwidth]{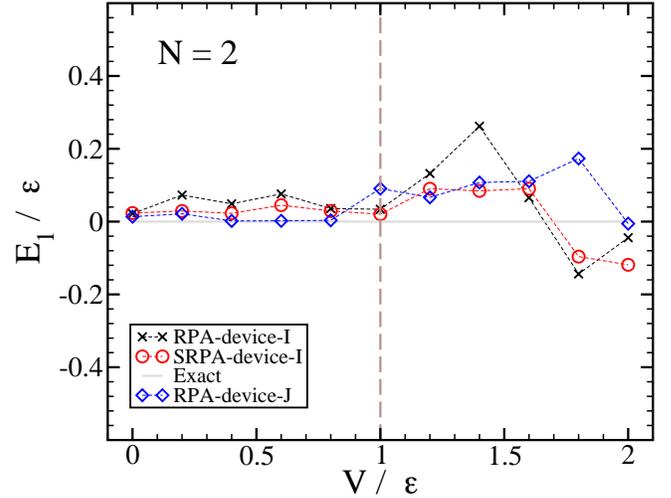}
	\cprotect\caption{First excited state energy ($\Bar{E}_1=E_1/\epsilon$) as a function of the interaction strength($v=V/\epsilon$) for $N=2$ particles. We compare the computational accuracy with configuration complexity $\alpha=1$ (RPA) for \textbf{I} and \textbf{J} encoding schemes, and $\alpha=2$ (SRPA) for the \textbf{I}-scheme.}
	\label{fig:Energy_E1_N2}
\end{figure}

For N=2, the energy spectrum as a function of the interaction strength is shown in Fig. \ref{fig:Energy_Spectrum_N2}. In the weak coupling regime, the qEOM solution for the excited state energies, for both encoding schemes, are relatively close to the exact solution. In the strong coupling regime, the \textbf{J}-scheme is slightly more accurate than the \textbf{I}-scheme, but they both deviate from the exact solution for a few points. To compare the effect of the configuration complexity ($\alpha$)  on the accuracy of the results, we consider the energy ($\Bar{E}_1$) of the first excited state as a function of the interaction strength. In the \textbf{I}-scheme, we compute $\Bar{E}_1$ using $\alpha=1$ (RPA-I) and $\alpha=2$ (SRPA-I), whereas in the \textbf{J}-scheme we only use $\alpha=1$ (RPA-J) since one cannot encode $2p2h$ configurations on one qubit. In the weak coupling regime, the \textbf{J}-scheme RPA result is relatively close to the exact solution. The \textbf{I}-scheme SRPA result is slightly more accurate than its RPA solution, but they are both less accurate than RPA in \textbf{J}-scheme. In the strong coupling regime, as shown in Fig. \ref{fig:Energy_E1_N2} and Table (\ref{table:Dev_Errors}), the \textbf{I}-scheme SRPA is as accurate as the \textbf{J}-scheme RPA, which are both more accurate than the \textbf{I}-scheme RPA. The simulation results for $\Bar{E}_1$ can be summarized as follows:

\begin{equation}\label{eq_E1_results}
\Bar{E}_1 \rightarrow \left\{\begin{array}{ll}
\text{RPA-J} > \text{SRPA-I} > \text{RPA-I}  & \text{for} \,\,\, v < 1 , \\
\text{SRPA-I} \approx \text{RPA-J} > \text{RPA-I}  & \text{for} \,\,\, v > 1 .
\end{array}\right.
\end{equation}
This shows that increasing the configuration complexity does improve the accuracy, as seen by the results for SRPA-I being slightly better than RPA-I for all values of the coupling strength. Although these results are not conclusive at this size of the system, they are sufficient to indicate that larger $\alpha$ leads to better accuracy. We also note, from the results of RPA-J being better than SRPA-I at small coupling strength, that a more efficient encoding scheme can reduce the degree of $\alpha$ required to achieve a certain accuracy. 

\begin{figure*}
	\centering
	\includegraphics[width=\linewidth]{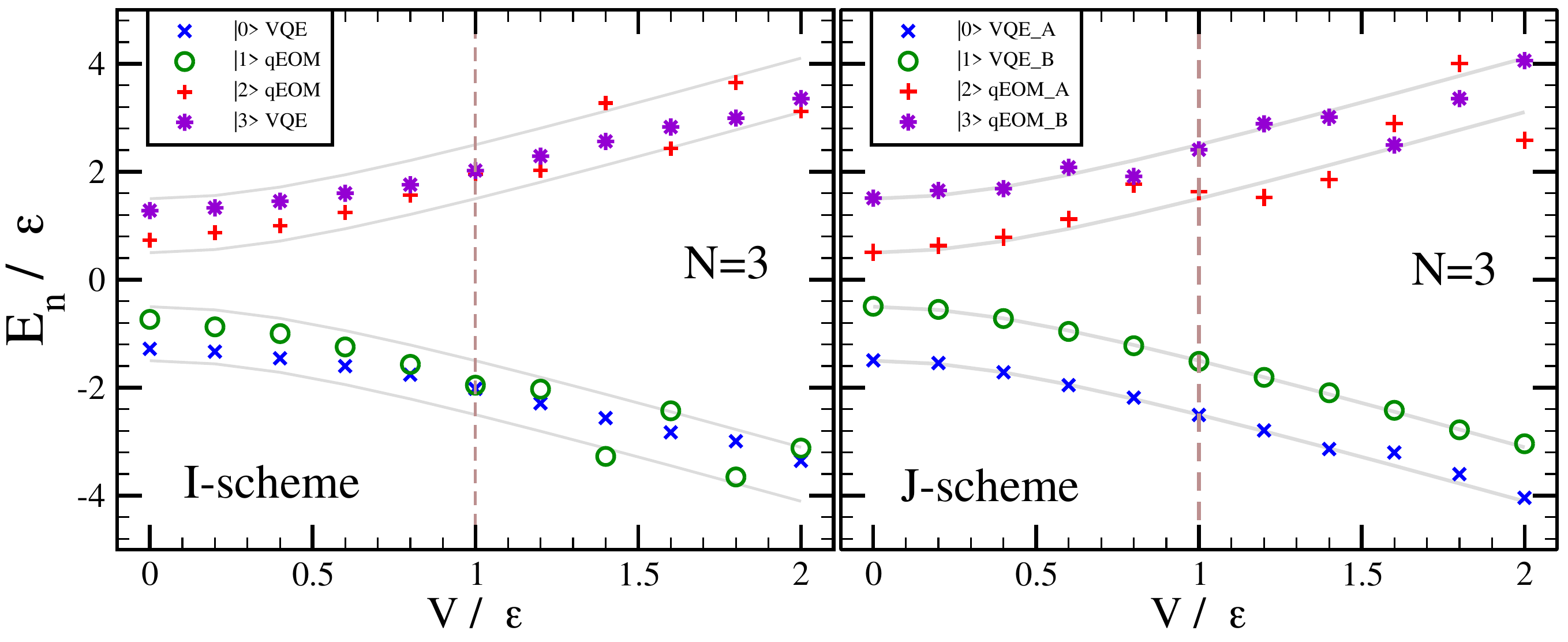}
	\cprotect\caption{Energy spectrum ($ \Bar{E}_n=E_n/\epsilon$) as a function of the interaction strength ($v=V/\epsilon$) for $N=3$ particles.}
	\label{fig:Energy_J_N3}
\end{figure*}

For N=3, the energy spectrum as a function of the  interaction strength is shown in Fig. \ref{fig:Energy_J_N3}. The results for the \textbf{I}- and \textbf{J}-schemes are displayed in the left and in the right panels, respectively. For the \textbf{J}-scheme plot, the legend post-factor labels of "A" and "B" denote block A and block B of the Hamiltonian, which are defined in Appendix \ref{appendix:encoding_schemes}. Using the symmetry of the LMG solutions when $W=0$, we construct the plot for the \textbf{I}-scheme as follows: $\Bar{E}_0$ (using VQE), $\Bar{E}_1 = -\Bar{E}_2$ , $\Bar{E}_2$ (using qEOM-SRPA), and $\Bar{E}_3 = -\Bar{E}_0$. Essentially, only $\Bar{E}_0$ and $\Bar{E}_2$ were computed, and $\Bar{E}_1$ and $\Bar{E}_3$ were found by reflecting $\Bar{E}_0$ and $\Bar{E}_2$ about the line $y=0$, respectively. For the \textbf{J}-scheme plot, recall that the Hamiltonian is split into two blocks, and each block uses one qubit (see Appendix \ref{appendix:encoding_schemes}) to compute $\Bar{E}_0, \Bar{E}_1$ (using VQE) and $\Bar{E}_2, \Bar{E}_3 $ (using qEOM-RPA).  In the weak coupling regime, results of the \textbf{J}-scheme are relatively close to the exact solution and significantly more accurate than the ones of the \textbf{I}-scheme. In the strong coupling regime, the \textbf{J}-scheme results are fairly close to the exact solution except for a few points with slightly larger errors. The \textbf{I}-scheme has noticeably larger errors with some points crossing the nearest energy level.\ 

In general, the results for the \textbf{I}-scheme are prone to more errors on current NISQ computers than those of our more efficient \textbf{J}-scheme because of the larger quantum resources (qubits and gates) required by the \textbf{I}-scheme. Another drawback of the \textbf{I}-scheme is that its Hamiltonian encoding introduces spurious energy solutions, that effectively increases the configuration complexity required to obtain the whole spectrum. This can be explicitly seen by rewriting the Hamiltonian of Eq. (\ref{eq_LMG_I_N3}) in the matrix form as
\begin{equation}\label{eq_LMG_I_N3_full}
	\Bar{H}^{(3)}_I = \frac{1}{2} \begin{pmatrix}
		3 & 0 & 0 & -2v & 0 & -2v & -2v & 0 \\
		0 & 1  & 0 & 0    & 0 & 0    & 0 & -2v \\
		 0 & 0  & 1 & 0  & 0 & 0 & 0  & -2v \\
	-2v  & 0 & 0 & -1   & 0 & 0 & 0 & 0  \\
	    0 & 0 & 0 & 0 & 1 & 0 & 0 & -2v \\
	   -2v & 0 & 0 & 0 & 0 & 1 & 0 & 0 \\
	   -2v & 0 & 0 & 0 & 0 & 0 & -1 & 0 \\
	   0 & -2v & -2v & 0 & 0 & -2v & 0  & -3
	\end{pmatrix}.
\end{equation}
Diagonalizing Eq. (\ref{eq_LMG_I_N3_full}), we get the energy solutions 
\begin{equation}
	\Bar{E}^{(3)} = \lbrace \pm \frac{1}{2}, \; \pm \sqrt{3v^2 + 1} - \frac{1}{2}, \; \frac{1}{2} \pm \sqrt{3v^2 + 1} \rbrace,
\end{equation}
where both of the extra energy solutions $\lbrace -\frac{1}{2}, \frac{1}{2} \rbrace$ have a two-fold degeneracy. This problem arises from encoding four active components of the ground state wave function onto three qubits which have eight possible states (see Appendix \ref{appendix:encoding_schemes}). Hence, the extra four non-active states give the additional non-physical energy solutions. This causes a hurdle for the qEOM as it will treat the extra non-active states as legitimate excited states, thus, the excitation operator will require a higher configuration complexity $\alpha > 3$ to get the complete energy spectrum. Note that for $N=3$, a comparison between configuration complexity similarly to one shown in Fig. \ref{fig:Energy_E1_N2} for the \textbf{I}-scheme is not meaningful since the result of qEOM with $\alpha = 1$ gives the unphysical solution $\Bar{E}_1 = \frac{1}{2}$. Although the \textbf{J}-scheme does not have this problem of spurious solutions for $N=3$, we cannot make the comparison between configuration complexities at this scale because only  $\alpha = 1$ (RPA) configuration is possible on one qubit. It is worth mentioning again that non-active states also appear in the \textbf{J}-encoding scheme, but they are much fewer than for the \textbf{I}-scheme and do not always add unphysical solutions. 

\begin{figure*}
	\centering
	\includegraphics[width=\linewidth]{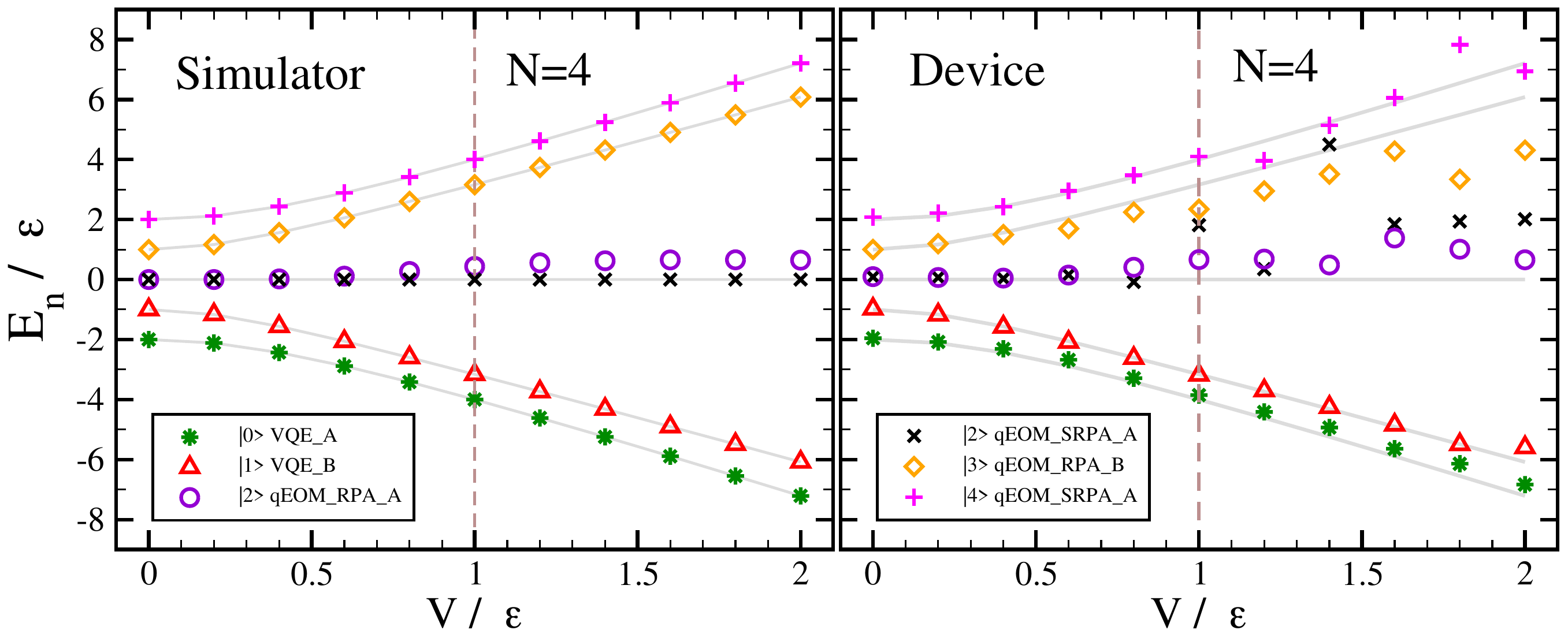}
	\cprotect\caption{Energy spectrum ($ \Bar{E}_n=E_n/\epsilon$) as a function of the interaction strength ($v=V/\epsilon$) for $N=4$ particles from a \verb|state_vector| simulator (left) and \verb|ibmq_quantum| computer (right). }
	\label{fig:Energy_J_N4}
\end{figure*}

For N=4, the energy spectrum as a function of the interaction strength is shown in Fig. \ref{fig:Energy_J_N4} where the legend post-factor labels of "A" and "B" denote block A with even $M$ values, and block B with odd $M$ values of the Hamiltonian, as shown in Sec. (\ref{sec:example_N4}). Diagonalizing the Hamiltonian in block A gives the energies $\Bar{E}_0$, $\Bar{E}_2$, and $\Bar{E}_4$ computed using VQE, qEOM-RPA/SRPA, and qEOM-SRPA, respectively. Similarly, for the block B we get the energies $\Bar{E}_1$ and $\Bar{E}_2$ computed using VQE and qEOM-RPA, respectively. First, we run the computation on a \verb|state_vector| simulator and find that for $\Bar{E}_2$ from block A Hamilitonian, the RPA solution significantly deviates from the exact solution as the interaction strength increases with average errors of about $10\%$ in the strong coupling region. In contrast, the SRPA solution stays relatively close to the exact solution with average errors of about $10^{-6}\%$ for all values of the coupling strength. This demonstrates explicitly that, in the absence of noise, an increase in configuration complexity $\alpha$ translates to increased accuracy of quantum many-body simulations within the EOM framework. This solidifies the interpretation of simulation results for $N=2$ when computing $\Bar{E}_1$ with RPA and SRPA, which is summarized in Eq. (\ref{eq_E1_results}). On a quantum device, the solutions for both $\Bar{E}_0$ and $\Bar{E}_1$ (computed using VQE) are fairly close to the exact ones in all coupling regimes. The qEOM results for the excited states $\Bar{E}_2, \Bar{E}_3$ and $\Bar{E}_4$ are fairly close to the exact ones in the weak coupling regime, but significantly deviate at the strong coupling. The average errors of the excited state energies found by qEOM in the strong coupling regime are larger than we expected considering the  VQE errors of the ground states they are computed from. This highlights the issue of a non-trivial error propagation in the qEOM algorithm, that is discussed in more detail in the Appendix of Ref. \cite{Ollitrault_2020}.

\section{\label{sec:summary} Summary and Outlook }

We simulated the excited states of the Lipkin model on a quantum computer using the Quantum Equation of Motion, which is an extension of the Variational Quantum Eigensolver. The goal was to find, within the equation of motion framework, how the configuration complexity ($\alpha$) of the many-body states correlates with the accuracy of the resulting spectra when simulated on a quantum computer. To achieve this objective, we first proposed a new efficient encoding scheme (the \textbf{J}-scheme) of the Lipkin model that exploits symmetries of the Hamiltonian and employs the Gray code to minimize the quantum resources needed for the simulation. Improving upon previously used encoding scheme \cite{cervia2021lipkin} (the \textbf{I}-scheme), our encoding scheme reduces the size of the Hilbert space from scaling as $\mathcal{O}(2^N)$ to $\mathcal{O}(\frac{N}{2} + 1)$ for a system of $N$ particles. This translates into reducing the number of qubits ($q_N$) from $q_N = N$ to the first integer that satisfies $q_N \geq \log_2\left( \frac{N}{2} + 1\right)$, and also makes the circuit depth shallower. We considered systems with $N=2, 3$ and $4$ particles, and run the simulations on IBM quantum computers and a \verb|state_vector| simulator. We compared simulations using our \textbf{J}-scheme and the \textbf{I}-scheme with configuration complexities $\alpha = 1$ (RPA) and $\alpha = 2$ (SRPA), for both weak ($v<1$) and strong ($v>1$) coupling regimes.  \\

On the simulator, for systems with $N=2$ and $3$ particles, we found no significant difference between the results of both encoding schemes and the exact solution in all coupling regimes with RPA and SRPA configuration complexities. By computing the average errors of the sampled points in each coupling regime, as shown in Table (\ref{table:Sim_Errors}), we found the average errors less than $10^{-5}\%$ for the aforementioned cases. However, as we increased the particle number to $N=4$, we observed an emerging difference between the RPA and SRPA solutions. Using the \textbf{J}-scheme, we saw that the RPA solution significantly deviates from the exact solution with errors that are more than five orders of magnitude larger than those of the SRPA solution. In the absence of noise, this clear difference between the RPA and SRPA solutions demonstrates that the configuration complexity directly impacts the accuracy of quantum many-body simulations, and by working with model Hamiltonians we can quantify how this scales with an increase in the interaction strength.\ 
On a quantum computer, we found that our \textbf{J}-scheme had significantly more accurate results than the \textbf{I}-scheme, and the difference becomes accentuated with an increase of any of the model parameters $\lbrace N, v, \alpha \rbrace$. For both encoding schemes, the simulations accumulated more errors in the strong coupling regime than in the weak coupling one, which is consistent with our intuition that quantum states that strongly interact should be more difficult to simulate than ones that weakly interact. For some cases, the excited state energy solution in the strong coupling regime appeared somewhat chaotic, which reveals one of the drawbacks of the qEOM. As discussed in the appendix of Ref. \cite{Ollitrault_2020}, it is non-trivial to predict how the errors from the ground state will propagate to the excited states in the presence of noise. \

Another drawback of the qEOM is that the matrix dimensions of the generalized eigenvalue equation (GEE), given in Eq. (\ref{eq_gev}), scales badly with the parameters $\left( N, \alpha \right)$. For $\alpha=2$, we found that the matrix dimensions of the GEE scales as $\mathcal{O}(N^2)$, thus we expect a general scaling (assuming no approximations made) of $\mathcal{O}(N^{\alpha})$ which is hardly manageable by classical computers for large $N$. A possible way to avoid the expensive diagonalization of a large GEE can be, for instance, to perform calculations on a spatial grid, introduce particle-vibration coupling or to use some form of the Finite Amplitude Method (FAM)  ~\cite{NakatsukasaInakuraYabana2007,Kortelainen2015,NiksicKraljTutisEtAl2013}. The latter leads to a series of differential equations that can be efficiently solved by numerical integration. Work in this area is currently in progress ~\cite{zhang2021many}. 

In this work we have demonstrated, using an exactly solvable quantum many-body model, that increasing the configuration complexity within the EOM framework increases the accuracy of our simulation. However, increasing $\alpha$ also increases the dimension of the matrices of the GEE to be solved on a classical computer. We also observed that an increase of the interaction strength produces a decrease in the accuracy of the simulation. As a way to combat these issues, we have proposed an efficient encoding scheme which i) minimizes the number of quantum resources required, thus reducing the errors in the strong coupling regime, ii) minimizes the configuration complexity $\alpha$ required to obtain accurate spectra for a given system, thus reducing the size of the GEE matrices. An example for the latter point is the case of $N=3$, where our scheme only required $\alpha=1$ to obtain the whole spectra instead of the theoretically exact $\alpha=N=3$. We also found that, as we increased the number of particles $N$, the accuracy of our simulations on a quantum computer declined due to an increase in the noise because an increase in $N$ essentially increases the effective coupling strength $\tilde{v} = (N-1) v/\epsilon$. Further work needs to be done to combat noise errors by employing error mitigation strategies, using better quality qubits and eventually employing quantum error correction in the near future. Our scheme and observations form a stepping stone towards developing quantum algorithms to achieve nuclear spectroscopic accuracy.

\begin{acknowledgments}

We thank Ben Hall and Kyle Wendt for valuable discussions on the Lipkin model and quantum simulations, respectively. MQH, YZ and EL acknowledges funding from US-NSF through the CAREER Grant PHY-1654379. DL acknowledges financial support from the CNRS through the 80Prime program and is part of the QC2I project. 
We appreciate cloud access of IBM quantum computers to run simulations for parts of this work. The views expressed are those of the
authors, and do not reflect the official policy or position of IBM.

\end{acknowledgments}

\appendix

\section{More on encoding schemes}\label{appendix:encoding_schemes} 

In this section, we give more examples of our efficient encoding scheme and associated ansatz circuit to simulate on a quantum computer.

\subsection{N=2}

Starting with Eq. (\ref{eq_LMG_J}), we consider a system with $N=2$ particles and two possible states $\ket{J, M} \rightarrow \ket{1, -1}, \ket{1, +1} $, so that the Hamiltonian is given by 
\begin{equation}\label{eq_LMG_J_2}
    \Bar{H}^{(2)}_J = a_0 \ket{0}\bra{0} + a_1 \ket{1}\bra{1} + b_0 \left[ \ket{0}\bra{1} + \ket{1}\bra{0} \right].
\end{equation}
We note that for this case the Standard Binary (SB) code and Gray code (GC) are identical. Applying the projection operators as described in Ref.~\cite{di2021improving}, we can rewrite Eq. (\ref{eq_LMG_J_2}) as
\begin{equation}\label{eq_LMG_J_2a}
\begin{split}
     \Bar{H}^{(2)}_J &= a_0 P^{(0)} + a_1 P^{(1)} + b_0 X \\
     &= \frac{1}{2} \left( a_0 + a_1 \right) + \frac{1}{2}\left( a_0 - a_1 \right) Z  + b_0 X,
\end{split}
\end{equation}
where $P^{(0)} = \frac{1}{2} \left( \mathbb{I} + Z \right) $ and $P^{(1)} = \frac{1}{2} \left( \mathbb{I} - Z \right)$ denote the projectors on the state $\ket{0}$ and $\ket{1}$, respectively. Using Eqs. (\ref{eq_ak}) and (\ref{eq_bk}), we find  
\begin{equation}
 a_0 = - a_1 = -1 \; \; \text{and} \; \; b_0 = - v \; ,
\end{equation}
so that the final Hamiltonian reads:
\begin{equation}\label{eq_LMG_J_2_final}
   \Bar{H}^{(2)}_J = -Z - v X = \begin{pmatrix}
   -1 & -v \\
   -v &  + 1
   \end{pmatrix},
\end{equation}
with the energy eigenvalues $\Bar{E}^{(2)} = \pm \sqrt{v^2 + 1}$. The third solution with energy $\Bar{E}^{(2)}=0$ is found from the second part of the full Hamiltonian containing the state $\ket{1,0}$. \\

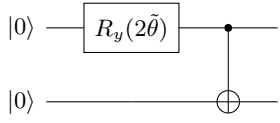
\begin{figure}[h!]
\centering
\begin{quantikz}[thin lines]
\lstick{$\ket{0}$} & \gate{R_y(2\tilde{\theta})}  & \ctrl{1} & \qw \\
\lstick{$\ket{0}$} & \qw & \targ{} & \qw
\end{quantikz}
\caption{I-scheme ansatz for N=2}
\label{fig:I_N2_ansatz}
\end{figure}

Using the encoding notation of Ref.~\cite{cervia2021lipkin}, the ground state wave function, in terms of the individual spin basis (\textbf{I}-scheme), is a superposition of two states  
\begin{equation}
\begin{split}
    \ket{\psi_I(\theta)} &= \sin{\theta}\ket{\uparrow \uparrow} - \cos{\theta}\ket{\downarrow \downarrow} \\
     &= \cos{\tilde{\theta}}\ket{00} + \sin{\tilde{\theta}}\ket{11},  
\end{split}
\end{equation}
where $ \tilde{\theta} = \theta -\frac{\pi}{2}$ and $\theta \in [0, \frac{\pi}{2})$. The associated parameterized quantum circuit is shown in Fig. \ref{fig:I_N2_ansatz}, where the optimal $\tilde{\theta}_0$ that minimizes $\bra{\psi_I(\tilde{\theta})} \Bar{H}^{(2)}_I \ket{\psi_I(\tilde{\theta})}$ is found using VQE. 

\begin{figure}[h!]
\centering
    \begin{quantikz}[thin lines]
    \lstick{$\ket{0}$} & \gate{R_y(2\phi)} & \qw 
    \end{quantikz}
\caption{J-scheme ansatz for N=2}
\label{fig:J_N2_ansatz}
\end{figure}
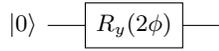

In our efficient \textbf{J}-scheme, the ground state wave function is given by
\begin{equation}
\begin{split}
    \ket{\psi_J(\phi)} &= \cos{\phi}\ket{1, -1} + \sin{\phi}\ket{1, +1} \\
 &= \cos{\phi}\ket{0} + \sin{\phi}\ket{1},  
\end{split}
\end{equation}
and the associated parameterized quantum circuit is shown in Fig. \ref{fig:J_N2_ansatz}, where $\phi \in [0, \frac{\pi}{2})$.

Note that, although both ans\"atze for \textbf{I}- and \textbf{J}-schemes have one parameter to optimize, the \textbf{J}-scheme is more efficient, because it uses less qubits and has a lower circuit depth than the \textbf{I}-scheme. 

\subsection{N=3}

For $N=3$, we have $J=\frac{3}{2}$, that corresponds to a total multiplet of 4 states, which decomposes into 2 disconnected sub blocks denoted by A and B:  
\begin{equation}\label{eq_N3_mapA}
\begin{aligned}
\ket{\frac{3}{2}, -\frac{3}{2}} & \equiv \ket{0}_A \qquad & \ket{\frac{3}{2}, +\frac{1}{2}}  & \equiv \ket{1}_A  \\
\ket{\frac{3}{2}, -\frac{1}{2}} & \equiv \ket{0}_B \qquad & \ket{\frac{3}{2}, +\frac{3}{2}}  & \equiv \ket{1}_B.
\end{aligned}  
\end{equation}
The Hamiltonian for both sets can be computed similarly to Eq. (\ref{eq_LMG_J_2}) 
\begin{equation}\label{eq_LMG_J_3}
\begin{split}
   \Bar{H}^{(3)}_J &= \frac{a_0}{2} \left( \mathbb{I} + Z \right) + \frac{a_1}{2} \left( \mathbb{I} - Z \right) + b_0 X \\
   &= \begin{pmatrix}
   a_0 & b_0 \\
   b_0 &  a_1
   \end{pmatrix},   
\end{split}
\end{equation}
where the coefficients are given by
\begin{equation}\label{eq_N3_coeff}
\begin{aligned}
  A: \quad & a_0 = -\frac{3}{2}  & a_1 = +\frac{1}{2} \qquad  b_0 = -v \sqrt{3}, \\
  B: \quad & a_0 = -\frac{1}{2}  & a_1 = +\frac{3}{2} \qquad  b_0 = -v \sqrt{3}. \\
\end{aligned}
\end{equation}
It is straightforward to verify that the combination of the two Hamiltonians generates the full energy spectrum 
\begin{equation}\label{eq_LMG_J_3_eng}
\Bar{E}^{(3)} =  
	\begin{cases}
		\Bar{E}_A = -\frac{1}{2} \pm \sqrt{3v^2 + 1}  \\
        \Bar{E}_B = +\frac{1}{2} \pm \sqrt{3v^2 + 1}  \\
	\end{cases}.       
\end{equation}

In the \textbf{J}-scheme, the ground state wave function is a superposition of four states ~\cite{cervia2021lipkin} given by
\begin{equation}\label{eq_psi_I_N3}
\begin{split}
    \ket{\psi_I(\theta)} &= \cos{\theta}\ket{\downarrow \downarrow \downarrow} -\frac{1}{\sqrt{3}} \sin{\theta}\left(\ket{\uparrow \uparrow \downarrow} + \ket{\uparrow  \downarrow \uparrow} + \ket{ \downarrow \uparrow   \uparrow} \right), \\
\end{split}
\end{equation}
where $\theta \in [0, \frac{\pi}{2})$. The associated parameterized quantum circuit is shown in Fig. \ref{fig:I_N3_ansatz}, where the auxiliary angles $\alpha$ and $\beta$ are defined to be

\begin{equation}
\begin{aligned}
&\alpha \equiv 2 \arccos \left(-\sqrt{\frac{2}{3}} \sin \theta\right), \\
&\beta \equiv-\frac{\pi}{4}-\arctan \left(\frac{\tan \theta}{\sqrt{3}}\right).
\end{aligned}
\end{equation}
Following the encoding notation of Ref. ~\cite{cervia2021lipkin}, we can write Eq. (\ref{eq_psi_I_N3}) as 
\begin{equation}
\begin{split}
    \ket{\psi_I(\theta)} &= \cos{\theta}\ket{111} -\frac{1}{\sqrt{3}} \sin{\theta}\left(\ket{001} + \ket{010} + \ket{100} \right). \\
\end{split}
\end{equation}
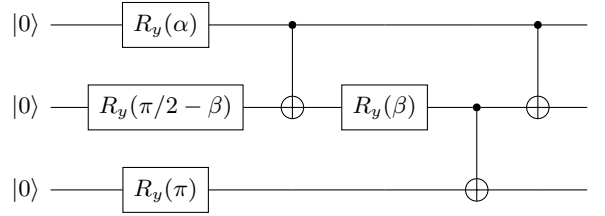
\begin{figure}[h!]
\centering
\begin{quantikz}[thin lines]
\lstick{$\ket{0}$} & \gate{R_y(\alpha)}        & \ctrl{1}  & \qw               & \qw      & \ctrl{1} & \qw \\
\lstick{$\ket{0}$} & \gate{R_y(\pi/2 - \beta)} & \targ{}   & \gate{R_y(\beta)} & \ctrl{1} & \targ{}  & \qw \\
\lstick{$\ket{0}$} & \gate{R_y(\pi)}           & \qw       & \qw               & \targ{}  & \qw      & \qw 
\end{quantikz}
\caption{I-scheme ansatz for N=3.}
\label{fig:I_N3_ansatz}
\end{figure}

\begin{figure}[h!]
\centering
    \begin{quantikz}[thin lines]
    \lstick{$\ket{0}$} & \gate{R_y(2\phi)} & \qw 
    \end{quantikz}
\caption{J-scheme ansatz for N=3.}
\label{fig:J_N3_ansatz}
\end{figure}
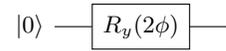

In the \textbf{J}-scheme, the wave function can be split into two blocks 
\begin{equation}
\ket{\psi_J(\phi)} =
\begin{cases}
\cos{\phi_A}\ket{\frac{3}{2}, -\frac{3}{2}} + \sin{\phi_A}\ket{\frac{3}{2}, +\frac{1}{2}} \\
\cos{\phi_B}\ket{\frac{3}{2}, -\frac{1}{2}} + \sin{\phi_B}\ket{\frac{3}{2}, +\frac{3}{2}} 
\end{cases},
\end{equation}
which can be solved independently for $\phi_A$ and $\phi_B$ using the quantum circuit shown in Fig. \ref{fig:J_N3_ansatz}, where $\phi_A = \phi_B + 2\pi r$ for $r \in \mathbb{Z}$.
Comparing the resources to solve the LMG model for $N=3$, we note that the \textbf{I}-scheme uses three qubits and seven gates, whereas the \textbf{J}-scheme requires only one qubit and one gate.

\section{qEOM matrices for $\alpha = 1$}\label{appendix:rpa_commutators} 

In this section we give the analytical expressions for the GEE matrices which enter Eq. (\ref{eq_gev}). In Sec. (\ref{sec:qEOM}), we evaluated the matrices $\mathcal{D}$ and $\mathcal{C}$ for $\alpha = 1$ to be

\begin{equation}
    \mathcal{D}_{mi^\prime kj^\prime} = - \braket{ \left[  \hat{a}^\dagger_{i^\prime} \hat{a}_{m}  , \hat{a}^\dagger_{j^\prime} \hat{a}_{k}   \right] } = 0, 
\end{equation}

\begin{equation}\label{eq_C1}
	\begin{split}
		\mathcal{C}_{mi^\prime kj^\prime} &=  \braket{ \left[  \hat{a}^\dagger_{i^\prime} \hat{a}_{m}  , \hat{a}^\dagger_{k} \hat{a}_{j^\prime }   \right] } \\
		&= \braket{ \left( \delta_{mk}  \hat{a}^\dagger_{i^\prime} \hat{a}_{j^\prime} - \delta_{i^\prime j^\prime}  \hat{a}^\dagger_{k} \hat{a}_{m}  \right) }. 
	\end{split}
\end{equation}

The evaluation of the remaining $\mathcal{A}$ and $\mathcal{B}$ matrices is more elaborate. First, we define the Hamiltonian to be used by setting $w=0$ into Eq. (\ref{eq_lmg_FS}) to get 

\begin{equation}
    \begin{split}
        \hat{H}_F &=  \frac{\epsilon}{2} \left(  \sum_{s^\prime=0}^{N-1} \hat{a}^\dagger_{s^\prime} \hat{a}_{s^\prime}  - \sum_{s=N}^{2N-1} \hat{a}^\dagger_{s} \hat{a}_{s}   \right)  \\
        & -\frac{V}{2}  \sum_{q,r} \sum_{q^\prime,r^\prime} \left( \hat{a}^\dagger_{q} \hat{a}^\dagger_{r} \hat{a}_{r^\prime} \hat{a}_{q^\prime}  + \hat{a}^\dagger_{q^\prime} \hat{a}^\dagger_{r^\prime} \hat{a}_{r} \hat{a}_{q}  \right) \Delta^{q,q^\prime}_{r,r\prime}.
    \end{split}
\end{equation}

Starting with the evaluation of matrix $\mathcal{A}$ given by
\begin{equation}\label{eq_Adef}
	\mathcal{A}_{mi^\prime kj^\prime} =  \braket{ \left[  \hat{a}^\dagger_{i^\prime} \hat{a}_{m}  , \left[ \hat{H}_F,  \hat{a}^\dagger_{k} \hat{a}_{j^\prime } \right]   \right] },
\end{equation}
we first compute the commutator 
\begin{equation}
	\left[  \hat{H}_0  , \hat{a}^\dagger_{k} \hat{a}_{j^\prime}   \right] = \frac{\epsilon}{2} \left( \sum_{s}\left[   \hat{a}^\dagger_{s} \hat{a}_{s} , \hat{a}^\dagger_{k} \hat{a}_{j^\prime}  \right]  - \sum_{s^\prime}\left[   \hat{a}^\dagger_{s^\prime} \hat{a}_{s^\prime} , \hat{a}^\dagger_{k} \hat{a}_{j^\prime}  \right] \right).
	\label{H0ph_comm} 
\end{equation}
The two commutators inside the sum can be evaluated as
\begin{eqnarray}
 & \left[   \hat{a}^\dagger_{s} \hat{a}_{s} , \hat{a}^\dagger_{k} \hat{a}_{j^\prime}  \right]  = \hat{a}^\dagger_{s} \hat{a}_{s} \hat{a}^\dagger_{k} \hat{a}_{j^\prime} -  \hat{a}^\dagger_{k} \hat{a}_{j^\prime} \hat{a}^\dagger_{s} \hat{a}_{s} \nonumber \\
 & = \hat{a}^\dagger_{s} \left( \delta_{sk} -  \hat{a}^\dagger_{k} \hat{a}_{s}  \right)   \hat{a}_{j^\prime} + \hat{a}^\dagger_{k} \hat{a}^\dagger_{s} \hat{a}_{j^\prime}  \hat{a}_{s} 
 = \delta_{sk} \hat{a}^\dagger_{s} \hat{a}_{j^\prime},\\
	& \left[   \hat{a}^\dagger_{s^\prime} \hat{a}_{s^\prime} , \hat{a}^\dagger_{k} \hat{a}_{j^\prime}  \right]  =  \hat{a}^\dagger_{s^\prime} \hat{a}_{s^\prime}  \hat{a}^\dagger_{k} \hat{a}_{j^\prime}  -  \hat{a}^\dagger_{k} \hat{a}_{j^\prime} \hat{a}^\dagger_{s^\prime} \hat{a}_{s^\prime} 
	\nonumber \\
	& = \hat{a}^\dagger_{s^\prime}   \hat{a}^\dagger_{k}  \hat{a}_{j^\prime} \hat{a}_{s^\prime}  -   \hat{a}^\dagger_{k} \left( \delta_{s^\prime j^\prime} -  \hat{a}^\dagger_{s^\prime} \hat{a}_{j^\prime}  \right) \hat{a}_{s^\prime} 
	\nonumber \\
&= -\delta_{s^\prime j^\prime} \hat{a}^\dagger_{k} \hat{a}_{s^\prime}.
\end{eqnarray}

Thus, the commutator of Eq. (\ref{H0ph_comm}) reads:
\begin{eqnarray}
	\left[  \hat{H}_0  , \hat{a}^\dagger_{k} \hat{a}_{j^\prime}   \right] &=& \frac{\epsilon}{2} \left( \sum_{s} \delta_{sk} \hat{a}^\dagger_{s} \hat{a}_{j^\prime}  - \sum_{s^\prime} \left(  -\delta_{s^\prime j^\prime} \hat{a}^\dagger_{k} \hat{a}_{s^\prime} \right) \right) \nonumber \\
	&=& \epsilon \hat{a}^\dagger_{k} \hat{a}_{j^\prime}. 
\end{eqnarray}
To evaluate the commutator associated with $\hat{H}_v$ given by
\begin{equation}
	\left[  \hat{H}_v  , \hat{a}^\dagger_{k} \hat{a}_{j^\prime}   \right] = -\frac{v}{2} \sum_{q,r} \sum_{q^\prime,r^\prime}   \left[ \hat{a}^\dagger_{q} \hat{a}^\dagger_{r} \hat{a}_{r^\prime} \hat{a}_{q^\prime} + \hat{a}^\dagger_{q^\prime} \hat{a}^\dagger_{r^\prime} \hat{a}_{r} \hat{a}_{q} ,  \hat{a}^\dagger_{k} \hat{a}_{j^\prime}   \right],  
\end{equation}
where the constraint $\Delta^{q,q^\prime}_{r,r\prime}$ will be inserted at the end of the calculation, we can simplify the commutators in the sum as follows:
\begin{eqnarray}
 & \left[ \hat{a}^\dagger_{q} \hat{a}^\dagger_{r} \hat{a}_{r^\prime} \hat{a}_{q^\prime}, \hat{a}^\dagger_{k} \hat{a}_{j^\prime}   \right]  =  \hat{a}^\dagger_{q} \hat{a}^\dagger_{r} \hat{a}_{r^\prime} \hat{a}_{q^\prime} \hat{a}^\dagger_{k} \hat{a}_{j^\prime}     -  \hat{a}^\dagger_{k} \hat{a}_{j^\prime}   \hat{a}^\dagger_{q} \hat{a}^\dagger_{r} \hat{a}_{r^\prime} \hat{a}_{q^\prime} \nonumber \\
 & = \hat{a}^\dagger_{k} \hat{a}^\dagger_{q} \hat{a}^\dagger_{r} \hat{a}_{r^\prime} \hat{a}_{q^\prime} \hat{a}_{j^\prime} - \hat{a}^\dagger_{k}  \hat{a}^\dagger_{q} \hat{a}^\dagger_{r} \hat{a}_{r^\prime} \hat{a}_{q^\prime} \hat{a}_{j^\prime} = 0. \\
 & \left[ \hat{a}^\dagger_{q^\prime} \hat{a}^\dagger_{r^\prime} \hat{a}_{r} \hat{a}_{q}, \hat{a}^\dagger_{k} \hat{a}_{j^\prime}   \right]  = \hat{a}^\dagger_{q^\prime} \hat{a}^\dagger_{r^\prime} \hat{a}_{r} \hat{a}_{q} \hat{a}^\dagger_{k} \hat{a}_{j^\prime}   - \hat{a}^\dagger_{k} \hat{a}_{j^\prime}  \hat{a}^\dagger_{q^\prime} \hat{a}^\dagger_{r^\prime} \hat{a}_{r} \hat{a}_{q} 
 \nonumber \\
 & = \hat{a}^\dagger_{q^\prime} \hat{a}^\dagger_{r^\prime} \left( \delta_{kq} \hat{a}_{r}   - \delta_{kr}\hat{a}_{q}  \right) \hat{a}_{j^\prime} + \hat{a}^\dagger_{k} \left( \delta_{j^\prime r^\prime} \hat{a}^\dagger_{q^\prime}  -  \delta_{j^\prime q^\prime}  \hat{a}^\dagger_{r^\prime}   \right) \hat{a}_{r} \hat{a}_{q}.\nonumber\\
\end{eqnarray}
For simplicity, we define
\begin{eqnarray}
    &\hat{f}_1 = \hat{a}^\dagger_{q^\prime} \hat{a}^\dagger_{r^\prime} \left( \delta_{kq} \hat{a}_{r}   - \delta_{kr}\hat{a}_{q}  \right) \hat{a}_{j^\prime},\\
   &\hat{f}_2 = \hat{a}^\dagger_{k} \left( \delta_{j^\prime r^\prime} \hat{a}^\dagger_{q^\prime}  -  \delta_{j^\prime q^\prime}  \hat{a}^\dagger_{r^\prime}   \right) \hat{a}_{r} \hat{a}_{q}. 
\end{eqnarray}

The double commutator of Eq. (\ref{eq_Adef}) consists of the two terms: 
\begin{equation}
\begin{split}
\left[  \hat{a}^\dagger_{i^\prime} \hat{a}_{m}  , \left[ \tilde{H},  \hat{a}^\dagger_{k} \hat{a}_{j^\prime } \right]   \right] &=  \left[  \hat{a}^\dagger_{i^\prime} \hat{a}_{m}  , \left[ \hat{H}_0,  \hat{a}^\dagger_{k} \hat{a}_{j^\prime } \right]   \right] \\
&+ \left[  \hat{a}^\dagger_{i^\prime} \hat{a}_{m}  , \left[ \hat{H}_v,  \hat{a}^\dagger_{k} \hat{a}_{j^\prime } \right]   \right].    
\end{split}
\end{equation}
The first commutator reads:
\begin{equation}
\begin{split}
	  & \left[  \hat{a}^\dagger_{i^\prime} \hat{a}_{m}  , \left[ \hat{H}_0,  \hat{a}^\dagger_{k} \hat{a}_{j^\prime } \right]   \right] = \left[ \hat{a}^\dagger_{i^\prime} \hat{a}_{m}  , \epsilon \hat{a}^\dagger_{k} \hat{a}_{j^\prime}  \right]\\
	 &= \epsilon \left( \delta_{mk}  \hat{a}^\dagger_{i^\prime} \hat{a}_{j^\prime} - \delta_{i^\prime j^\prime}  \hat{a}^\dagger_{k} \hat{a}_{m}    \right),
\end{split}
\end{equation}
while the second one gives:
\begin{equation}
	\left[  \hat{a}^\dagger_{i^\prime} \hat{a}_{m}  , \left[ \hat{H}_v,  \hat{a}^\dagger_{k} \hat{a}_{j^\prime } \right]   \right] = -\frac{v}{2} \sum_{q,r} \sum_{q^\prime,r^\prime} \left[  \hat{a}^\dagger_{i^\prime} \hat{a}_{m} , \hat{f}_1 +  \hat{f}_2    \right].
\end{equation}
Furthermore, we have:
\begin{equation}
	\left[  \hat{a}^\dagger_{i^\prime} \hat{a}_{m} , \hat{f}_1 \right] = - \delta_{i^\prime j^\prime} \hat{a}^\dagger_{q^\prime} \hat{a}^\dagger_{r^\prime} \left( \delta_{kq} \hat{a}_{r}   - \delta_{kr}\hat{a}_{q}  \right) \hat{a}_{m}
\end{equation}
and
\begin{equation}
	\left[  \hat{a}^\dagger_{i^\prime} \hat{a}_{m} , \hat{f}_2 \right] = \delta_{mk} \hat{a}^\dagger_{i^\prime} \left( \delta_{j^\prime r^\prime} \hat{a}^\dagger_{q^\prime}  -  \delta_{j^\prime q^\prime}  \hat{a}^\dagger_{r^\prime}   \right) \hat{a}_{r} \hat{a}_{q}. 
\end{equation}
Therefore, inserting all the pertinent terms into Eq. (\ref{eq_Adef}) we get 
\begin{equation}\label{eq_A}
	\begin{split}
		\mathcal{A}_{mi^\prime kj^\prime} &=  \braket{ \left[  \hat{a}^\dagger_{i^\prime} \hat{a}_{m}  , \left[ \tilde{H},  \hat{a}^\dagger_{k} \hat{a}_{j^\prime } \right]   \right] } \\
		&= \braket{   \epsilon \left(  \delta_{mk}  \hat{a}^\dagger_{i^\prime} \hat{a}_{j^\prime} - \delta_{i^\prime j^\prime}  \hat{a}^\dagger_{k} \hat{a}_{m}  \right) } \\
		&+\frac{v}{2} \sum_{q,r} \sum_{q^\prime,r^\prime} \braket{ \left[  \delta_{i^\prime j^\prime} \hat{a}^\dagger_{q^\prime} \hat{a}^\dagger_{r^\prime} \left( \delta_{kq} \hat{a}_{r}   - \delta_{kr} \hat{a}_{q}  \right) \hat{a}_{m}        \right] }\\
		&-\frac{v}{2} \sum_{q,r} \sum_{q^\prime,r^\prime} \braket{ \left[ \delta_{mk} \hat{a}^\dagger_{i^\prime} \left( \delta_{j^\prime r^\prime} \hat{a}^\dagger_{q^\prime}  -  \delta_{j^\prime q^\prime}  \hat{a}^\dagger_{r^\prime}   \right) \hat{a}_{r} \hat{a}_{q}    \right] },	\\
	\end{split}
\end{equation}
where the sums are constrained by the condition $\delta_{q+N,q^\prime } \delta_{r+N,r^\prime }$. Similarly, we can evaluate the $\mathcal{B}$ matrix defined as
\begin{equation}\label{eq_Bdef}
	\mathcal{B}_{mi^\prime kj^\prime} = - \braket{ \left[  \hat{a}^\dagger_{i^\prime} \hat{a}_{m}  , \left[ \tilde{H},  \hat{a}^\dagger_{ j^\prime } \hat{a}_{k } \right]   \right] },
\end{equation}
by fist evaluating the commutator associated with $\hat{H}_0$ as
\begin{align*}
	&\left[  \hat{H}_0  , \hat{a}^\dagger_{ j^\prime } \hat{a}_{k }   \right] = \frac{\epsilon}{2} \left( \sum_{s}\left[   \hat{a}^\dagger_{s} \hat{a}_{s} , \hat{a}^\dagger_{ j^\prime } \hat{a}_{k }  \right]  - \sum_{s^\prime}\left[   \hat{a}^\dagger_{s^\prime} \hat{a}_{s^\prime} , \hat{a}^\dagger_{ j^\prime } \hat{a}_{k } \right] \right)\\
	&=	 \frac{\epsilon}{2} \left( \sum_{s} -\delta_{sk}\hat{a}^\dagger_{ j^\prime } \hat{a}_{s} - \sum_{s^\prime} \delta_{s^\prime j^\prime}\hat{a}^\dagger_{ s^\prime } \hat{a}_{k} \right) = -\epsilon \hat{a}^\dagger_{ j^\prime } \hat{a}_{k},
\end{align*}
and the associated double commutator:
\begin{equation}
	\left[  \hat{a}^\dagger_{i^\prime} \hat{a}_{m}  , \left[ \hat{H}_0,  \hat{a}^\dagger_{ j^\prime } \hat{a}_{k } \right]   \right] = \left[  \hat{a}^\dagger_{i^\prime} \hat{a}_{m} , -\epsilon  \hat{a}^\dagger_{ j^\prime } \hat{a}_{k}  \right] = 0.
\end{equation}
Thus, the matrix $\mathcal{B}$ only contains information about the interaction term.
The respective commutator reads:
\begin{equation}
	\left[  \hat{H}_v  ,  \hat{a}^\dagger_{ j^\prime } \hat{a}_{k }   \right] = -\frac{v}{2} \sum_{q,r} \sum_{q^\prime,r^\prime}   \left[ \hat{a}^\dagger_{q} \hat{a}^\dagger_{r} \hat{a}_{r^\prime} \hat{a}_{q^\prime} + \hat{a}^\dagger_{q^\prime} \hat{a}^\dagger_{r^\prime} \hat{a}_{r} \hat{a}_{q} ,  \hat{a}^\dagger_{ j^\prime } \hat{a}_{k }   \right].  
\end{equation} 
Evaluating the first commutator in the sum leads to:
\begin{eqnarray}
	\left[ \hat{a}^\dagger_{q} \hat{a}^\dagger_{r} \hat{a}_{r^\prime} \hat{a}_{q^\prime} ,  \hat{a}^\dagger_{ j^\prime } \hat{a}_{k }   \right] = \hat{a}^\dagger_{q} \hat{a}^\dagger_{r} \hat{a}_{r^\prime} \hat{a}_{q^\prime} \hat{a}^\dagger_{ j^\prime } \hat{a}_{k }   - \hat{a}^\dagger_{ j^\prime } \hat{a}_{k } \hat{a}^\dagger_{q} \hat{a}^\dagger_{r} \hat{a}_{r^\prime} \hat{a}_{q^\prime}   \nonumber \\
	= \hat{a}^\dagger_{q} \hat{a}^\dagger_{r}  \left( \delta_{j^\prime q^\prime}\hat{a}_{r^\prime}  -   \delta_{j^\prime r^\prime}\hat{a}_{q^\prime}  \right) \hat{a}_{k } 
	+ \hat{a}^\dagger_{ j^\prime } \left( \delta_{kr}  \hat{a}^\dagger_{q}  - \delta_{kq} \hat{a}^\dagger_{r} \right) \hat{a}_{r^\prime} \hat{a}_{q^\prime}. \nonumber \\ 
\end{eqnarray}
Again, for simplicity we define
\begin{eqnarray}
   &\hat{g}_1 = \hat{a}^\dagger_{q} \hat{a}^\dagger_{r}  \left( \delta_{j^\prime q^\prime}\hat{a}_{r^\prime}  -   \delta_{j^\prime r^\prime}\hat{a}_{q^\prime}  \right) \hat{a}_{k } \\
&\hat{g}_2 = \hat{a}^\dagger_{ j^\prime } \left( \delta_{kr}  \hat{a}^\dagger_{q}  - \delta_{kq} \hat{a}^\dagger_{r} \right) \hat{a}_{r^\prime} \hat{a}_{q^\prime},
\end{eqnarray}
so that the second commutator gives 
\begin{equation}
\begin{split}
	 \left[  \hat{a}^\dagger_{q^\prime} \hat{a}^\dagger_{r^\prime} \hat{a}_{r} \hat{a}_{q} ,  \hat{a}^\dagger_{ j^\prime } \hat{a}_{k }   \right] =  \hat{a}^\dagger_{q^\prime} \hat{a}^\dagger_{r^\prime} \hat{a}_{r} \hat{a}_{q} \hat{a}^\dagger_{ j^\prime } \hat{a}_{k }     -  \hat{a}^\dagger_{ j^\prime } \hat{a}_{k } \hat{a}^\dagger_{q^\prime} \hat{a}^\dagger_{r^\prime} \hat{a}_{r} \hat{a}_{q}
	 \\
	 =\hat{a}^\dagger_{ j^\prime } \hat{a}^\dagger_{q^\prime} \hat{a}^\dagger_{r^\prime} \hat{a}_{r} \hat{a}_{q}  \hat{a}_{k }  - \hat{a}^\dagger_{ j^\prime } \hat{a}_{k } \hat{a}^\dagger_{q^\prime} \hat{a}^\dagger_{r^\prime} \hat{a}_{r} \hat{a}_{q} = 0,
	 \end{split}
\end{equation}
and the double commutator thus  reads: 
\begin{equation}
	\left[  \hat{a}^\dagger_{i^\prime} \hat{a}_{m}  , \left[ \hat{H}_v,  \hat{a}^\dagger_{ j^\prime } \hat{a}_{k } \right]   \right] = - \frac{v}{2} \sum_{q,r} \sum_{q^\prime,r^\prime} \left[ \hat{a}^\dagger_{i^\prime} \hat{a}_{m} , \hat{g}_1 + \hat{g}_2  \right].
	\label{dcHv}
\end{equation}
The first and the second terms of Eq. (\ref{dcHv}) become, respectively,
\begin{eqnarray}
\left[ \hat{a}^\dagger_{i^\prime} \hat{a}_{m} , \hat{g}_1  \right] = 	\hat{a}^\dagger_{i^\prime} \hat{a}_{m}  \hat{a}^\dagger_{q} \hat{a}^\dagger_{r}  \left( \hat{a}_{r^\prime}  -   \hat{a}_{q^\prime}  \right) \hat{a}_{k }  \nonumber \\
-  \hat{a}^\dagger_{q} \hat{a}^\dagger_{r}  \left( \hat{a}_{r^\prime}  -   \hat{a}_{q^\prime}  \right) \hat{a}_{k }  \hat{a}^\dagger_{i^\prime} \hat{a}_{m}\nonumber \\
 = \hat{a}^\dagger_{i^\prime}  \left( \delta_{mq}\hat{a}^\dagger_{r}  - \delta_{mr} \hat{a}^\dagger_{q}   \right)  \left( \hat{a}_{r^\prime}  -   \hat{a}_{q^\prime}  \right) \hat{a}_{k } \nonumber \\
+ \left(  \delta_{i^\prime q^\prime} - \delta_{i^\prime r^\prime} \right) \hat{a}^\dagger_{q} \hat{a}^\dagger_{r} \hat{a}_{m}\hat{a}_{k },\nonumber \\
\end{eqnarray}
and the second term
\begin{eqnarray}
	\left[ \hat{a}^\dagger_{i^\prime} \hat{a}_{m} , \hat{g}_2  \right] = \hat{a}^\dagger_{i^\prime} \hat{a}_{m}  \hat{a}^\dagger_{ j^\prime } \left(  \hat{a}^\dagger_{q}  -  \hat{a}^\dagger_{r} \right) \hat{a}_{r^\prime} \hat{a}_{q^\prime}  \nonumber \\
	-   \hat{a}^\dagger_{ j^\prime } \left(  \hat{a}^\dagger_{q}  -  \hat{a}^\dagger_{r} \right) \hat{a}_{r^\prime} \hat{a}_{q^\prime}  \hat{a}^\dagger_{i^\prime} \hat{a}_{m}\nonumber \\
	 =  \left(  \delta_{mr} -  \delta_{mq}  \right) \hat{a}^\dagger_{i^\prime} \hat{a}^\dagger_{ j^\prime } \hat{a}_{r^\prime} \hat{a}_{q^\prime} \nonumber \\
	+ \hat{a}^\dagger_{ j^\prime } \left(  \hat{a}^\dagger_{q}  -  \hat{a}^\dagger_{r} \right)   \left( \delta_{i^\prime r^\prime} \hat{a}_{q^\prime} -     \delta_{i^\prime q^\prime} \hat{a}_{r^\prime} \right)  \hat{a}_{m}.\nonumber\\
\end{eqnarray}
Therefore, inserting all the pertinent terms into Eq. (\ref{eq_Bdef}), we get
\begin{equation}\label{eq_B}
	\begin{split}
		\mathcal{B}_{mi^\prime kj^\prime} &= -  \braket{ \left[  \hat{a}^\dagger_{i^\prime} \hat{a}_{m}  , \left[ \tilde{H},  \hat{a}^\dagger_{ j^\prime } \hat{a}_{k } \right]   \right] } \\
		&= \frac{v}{2} \sum_{q,r} \sum_{q^\prime,r^\prime}  \lbrace \braket{  \left(  \delta_{mr} -  \delta_{mq}  \right) \hat{a}^\dagger_{i^\prime} \hat{a}^\dagger_{ j^\prime } \hat{a}_{r^\prime} \hat{a}_{q^\prime} } \\
		&+ \braket{ \left(  \delta_{i^\prime q^\prime} - \delta_{i^\prime r^\prime} \right) \hat{a}^\dagger_{q} \hat{a}^\dagger_{r} \hat{a}_{m}\hat{a}_{k } } \\
		&+ \braket{ \hat{a}^\dagger_{ j^\prime } \left( \delta_{kr} \hat{a}^\dagger_{q}  -  \delta_{kq} \hat{a}^\dagger_{r} \right)   \left( \delta_{i^\prime r^\prime} \hat{a}_{q^\prime} -     \delta_{i^\prime q^\prime} \hat{a}_{r^\prime} \right)  \hat{a}_{m} }	\\
		& 	+ \braket{ \hat{a}^\dagger_{i^\prime}  \left( \delta_{mq}\hat{a}^\dagger_{r}  - \delta_{mr} \hat{a}^\dagger_{q}   \right)  \left( \delta_{j^\prime q^\prime} \hat{a}_{r^\prime}  - \delta_{j^\prime r^\prime}  \hat{a}_{q^\prime}  \right) \hat{a}_{k }  } \rbrace ,
	\end{split}
\end{equation}
where the summation constraint $ \Delta^{q,q^\prime}_{r,r\prime}$ is not explicitly written for readability, but must be included in the terms associated with the $v$-scattering for both $\mathcal{A}$ and $\mathcal{B}$ matrices. 


\section{Error Analysis}\label{appendix:av_errors} 
In this section we give a pertinent analysis of the errors of the quantum equation of motion algorithm for the Lipkin model when implemented on current noisy quantum computers with calibration data given by Table \ref{table:av_calibration}. 

\begin{center}
\captionof{table}{Range of average calibration data for the IBM quantum machines used for the simulations in this work. \label{table:av_calibration}}
\begin{tabular}{lcr}
\hline \hline 
Parameter & lower & upper \\
\hline
Av. $T_{1}(\mu \mathrm{s})$ & $93.27$ & $133.38$ \\
Av. $T_{2}(\mu \mathrm{s})$ & $56.43$ & $137.67$ \\
Av. CNOT Error & 7.919e-3 & 3.29e-1 \\
Av. Readout Error & 2.858e-2 & 3.208e-2 \\
\hline \hline
\end{tabular}
\end{center}

For analyzing the accuracy of our methods, we compute the average percentage error of each coupling regime given by

\begin{equation}
\label{Delta_err}
    \Delta_n \equiv 
    \begin{cases}
      \sum_{i}^{p} \frac{m_n^{(i)}-\epsilon_0^{(i)}  }{\epsilon_n^{(i)}- \epsilon_0^{(i)}} \; \; \text{for} \; \; n>0 \\
    \sum_{i}^{p} \frac{m_0^{(i)}  }{\epsilon_0^{(i)}} \; \; \text{for} \; \; n=0. 
    \end{cases}
\end{equation}
Here $(p-i)$ is half the number of data points, and $m_n^{(i)}$ and $\epsilon_n^{(i)}$ are the $i^{th}$ measured and exact energy points for the $n^{th}$ energy level, respectively. For each coupling regime, Tables \ref{table:Sim_Errors} and \ref{table:Dev_Errors} show a  comparison of the errors of VQE and qEOM implementations on the simulator and quantum device, respectively.

\begin{table*}[h]
\begin{tabular}{|c|c|c|c|c|c|c|c|c|c|c|c|c|c|c|}
\hline \multicolumn{15}{|c|}{ \textbf{Simulator} } \\
\hline \multirow{ 4}{*}{}  & & \multicolumn{6}{|c|}{ J-scheme } &  &  \multicolumn{6}{|c|}{ I-scheme } \\
\cline{3-15} & & \multicolumn{3}{|c|}{ Weak $(v<1)$} & \multicolumn{3}{|c|}{ Strong $(v>1)$} &  &  \multicolumn{3}{|c|}{ Weak $(v<1)$} & \multicolumn{3}{|c|}{ Strong $(v>1)$} \\
 \cline{3-15} & & \multirow{ 2}{*}{VQE}  & \multicolumn{2}{|c|}{ qEOM } & \multirow{ 2}{*}{VQE} & \multicolumn{2}{|c|}{ qEOM } &  &  \multirow{ 2}{*}{VQE} & \multicolumn{2}{|c|}{ qEOM } & \multirow{ 2}{*}{VQE} & \multicolumn{2}{|c|}{ qEOM } \\
\cline{4-5} \cline{7-8} \cline{11-12} \cline{14-15}  & & & RPA & SRPA & & RPA & SRPA &  &  & RPA & SRPA & & RPA & SRPA \\
\hline \multirow{ 3}{*}{$N=2$}  & $E_0$ & 1.37e-07 &  &  & 1.79e-07 &  &  & - & 1.37e-07 &  &  & 1.79e-07 &  &  \\
 \cline{2-15} &   $E_1$ &  & 3.06e-06 &  &  & 1.37-05 &  & - &  & 4.07e-05 & 4.07-05 & & 2.53e-05 & 2.53e-05 \\
\cline{2-15}  &  $E_2$ &  & 3.03e-06 &  &  & 1.37e-05 &  & - &  &  & 2.04e-05 &  &  & 1.27e-05 \\
\hline \multicolumn{15}{|c|}{} \\
\hline \multirow{ 4}{*}{$N=3$}  & $E_0$ & 1.41e-10 &  &  & 4.42e-13 &  &  & - & 5.05e-11 &  &  & 4.19e-11 &  &  \\
 \cline{2-15} &   $E_1$ & 4.33e-10 &  &  & 6.47e-12 &  &  & - &  &  & 4.48e-05 &  &  & 2.80e-04 \\
\cline{2-15}  &  $E_2$ &  & 1.59e-05 &  &  & 7.28e-06 &  & - &  &  & 4.48e-05 &  &  & 2.80e-04 \\
\cline{2-15}  &  $E_3$ &  & 3.65e-07  &  &  & 1.82e-05 &  & - & 5.05e-11 &  &  & 4.19e-11 &  &  \\
\hline \multicolumn{15}{|c|}{} \\
\hline \multirow{ 5}{*}{$N=4$}  & $E_0$ & 1.35e-11 &  &  & 3.95e-13 &  &  & - &  &  &  &  &  &  \\
 \cline{2-15} &   $E_1$ & 1.01e-11 &  & 4.88e-14 &  &  &  & - &  &  &  &  &  &  \\
\cline{2-15}  &  $E_2$ &  & \textbf{2.50e+00} & 1.30e-05  &  & \textbf{1.08e+01} & 1.00e-05 & - &  &  &  &  &  &  \\
\cline{2-15}  &  $E_3$ &  & 1.32e-05 &  &  & 1.79e-06 &  & - &  &  &  &  &  &  \\
\cline{2-15}  &  $E_4$ &  &  & 2.25e-06 &  &  & 2.42e-06 & - &  &  &  &  &  &  \\
\hline
\end{tabular}
\cprotect\caption{Summary of average percentage errors from \verb|state_vector| simulator. }
\label{table:Sim_Errors}
\end{table*}

\begin{table*}[h]
\begin{tabular}{|c|c|c|c|c|c|c|c|c|c|c|c|c|c|c|}
\hline \multicolumn{15}{|c|}{ \textbf{Device} } \\
\hline \multirow{ 4}{*}{}  & & \multicolumn{6}{|c|}{ J-scheme } &  &  \multicolumn{6}{|c|}{ I-scheme } \\
\cline{3-15} & & \multicolumn{3}{|c|}{ Weak $(v<1)$} & \multicolumn{3}{|c|}{ Strong $(v>1)$} &  &  \multicolumn{3}{|c|}{ Weak $(v<1)$} & \multicolumn{3}{|c|}{ Strong $(v>1)$} \\
 \cline{3-15} & & \multirow{ 2}{*}{VQE}  & \multicolumn{2}{|c|}{ qEOM } & \multirow{ 2}{*}{VQE} & \multicolumn{2}{|c|}{ qEOM } &  &  \multirow{ 2}{*}{VQE} & \multicolumn{2}{|c|}{ qEOM } & \multirow{ 2}{*}{VQE} & \multicolumn{2}{|c|}{ qEOM } \\
\cline{4-5} \cline{7-8} \cline{11-12} \cline{14-15}  & & & RPA & SRPA & & RPA & SRPA &  &  & RPA & SRPA & & RPA & SRPA \\
\hline \multirow{ 3}{*}{$N=2$}  & $E_0$ & 1.17e+00 &  &  & 2.52e+00 &  &  & - & 2.38e+00 &  &  & 6.16e+00 &  &  \\
 \cline{2-15} &   $E_1$ &  & 8.10e-01 &  &  & 5.01e+00 &  & - &  & 4.64e+00 & 2.69e+00 & & 6.75e+00 & 5.16e+00 \\
\cline{2-15}  &  $E_2$ &  & 5.82e-01 &  &  & 4.23e+00 &  & - &  &  & 1.50e+00 &  &  & 5.16e+00 \\
\hline \multicolumn{15}{|c|}{} \\
\hline \multirow{ 4}{*}{$N=3$}  & $E_0$ & 6.23e-01 &  &  & 2.86e+00 &  &  & - & 1.64e+01 &  &  & 1.87e+01 &  &  \\
 \cline{2-15} &   $E_1$ & 9.56e-01 &  &  & 2.59e+00 &  &  & - &  &  & 2.98e+01 &  &  & 3.97e+01 \\
\cline{2-15}  &  $E_2$ &  & 5.86e+00 &  &  & 9.00e+00 &  & - &  &  & 2.98e+01 &  &  & 3.97e+01 \\
\cline{2-15}  &  $E_3$ &  & 2.91e+00 &  &  & 4.63e+00 &  & - & 1.64e+01 &  &  & 1.87e+01 &  &  \\
\hline \multicolumn{15}{|c|}{} \\
\hline \multirow{ 5}{*}{$N=4$}  & $E_0$ & 3.79e+00 &  &  & 4.91e+00 &  &  & - &  &  &  &  &  &  \\
 \cline{2-15} &   $E_1$ & 4.05e+00 &  &  & 1.70e+00 &  &  & - &  &  &  &  &  &  \\
\cline{2-15}  &  $E_2$ &  & 5.44e+00 & 3.39e+00  &  & 1.44e+01 & 3.64e+01 & - &  &  &  &  &  &  \\
\cline{2-15}  &  $E_3$ &  & 3.14e+00 &  &  & 1.09e+01 &  & - &  &  &  &  &  &  \\
\cline{2-15}  &  $E_4$ &  &  & 1.32e+00 &  &  & 4.21e+00 & - &  &  &  &  &  &  \\
\hline
\end{tabular}
\cprotect\caption{Summary of average percentage errors from \verb|ibmq_quantum| computers.}
\label{table:Dev_Errors}
\end{table*}

\clearpage

\bibliography{references1}

\end{document}